\documentclass[12pt]{article}
\usepackage{amsmath}
\usepackage{graphicx}
\usepackage{enumerate}
\usepackage{natbib}
\usepackage{multirow}
\usepackage{longtable}
\usepackage{booktabs}
\usepackage{amssymb}
\usepackage{bm}
\usepackage{mathrsfs}
\usepackage{amsthm}
\usepackage{amsfonts}
\usepackage{subfigure}
\usepackage{floatrow}
\usepackage{epsfig,amssymb,latexsym,verbatim}
\usepackage{threeparttable}
\usepackage{epstopdf}
\usepackage{graphics}
\usepackage[english]{babel}
\usepackage[figuresright]{rotating}
\usepackage[dvipsnames]{xcolor}
\usepackage{color}
\usepackage{hyperref}

\usepackage{cases}
\usepackage{makecell}
\usepackage{caption}
\usepackage{filecontents}
\usepackage{algorithmic}
\usepackage{algorithm}
\usepackage{}

\newtheorem{theorem}{Theorem}[section]
\newtheorem{corollary}{Corollary}[section]

\newtheorem{proposition}{Proposition}[section]

\newtheorem{remark}{Remark}[section]

\newtheorem{assumption}{Assumption}[section]

\usepackage{url} % not crucial - just used below for the URL 

\newcommand{\blind}{0}
\usepackage{xr}
\def\Var{\mbox{Var}}
\def\Cov{\mbox{Cov}}

\allowdisplaybreaks

\renewcommand{\hat}{\widehat}
\renewcommand{\tilde}{\widetilde}
\newcommand{\bX}{\boldsymbol{X}}

\newcommand{\tr}{\mathrm{tr}}
\begin{document}

	\def\spacingset#1{\renewcommand{\baselinestretch}%
		{#1}\small\normalsize} \spacingset{1}

	%%%%%%%%%%%%%%%%%%%%%%%%%%%%%%%%%%%%%%%%%%%%%%%%%%%%%%%%%%%%%%%%%%%%%%%%%%%%%%
	\date{}
	\if0\blind
		\title{\bf  Adaptive adequacy testing of high-dimensional factor-augmented regression model}
        \author{Yanmei Shi$^1$, Leheng Cai$^2$, Xu Guo$^1$\thanks{Corresponding author: Xu Guo. Email address: xustat12@bnu.edu.cn.}, and Shurong Zheng$^3$
		\\~\\
		{\small \it $^{1}$  School of Statistics, Beijing Normal University, Beijing, China}
		\\{\small \it $^2$ Department of Statistics and Data Science,   Tsinghua University, Beijing, China}\\{\small \it $^{3}$ School of Mathematics and Statistics, Northeast Normal University, Changchun, China}
 }
%		\author{Yanmei Shi\footnotemark[1], Leheng Cai\footnotemark[2],   Xu Guo\footnotemark[1] \footnotemark[3], Shurong Zheng\footnotemark[4]  \hspace{.2cm}}
%		\maketitle
%		\renewcommand{\thefootnote}{\fnsymbol{footnote}}
%		\footnotetext[1]{School of Statistics, Beijing Normal University, Beijing, China}
%		\footnotetext[2]{ Department of Statistics and Data Science,   Tsinghua University, Beijing, China}
%		\footnotetext[3]{Corresponding Author: Xu Guo. Email address: xustat12@bnu.edu.cn.}
%        \footnotetext[4]{School of Mathematics and Statistics, Northeast Normal University, Changchun, China}
 	\date{}
	%\date{\today}

	\maketitle
	
	\if1\blind
	{
		\bigskip
		\bigskip
		\bigskip
		\begin{center}
			{\LARGE\bf Adaptive adequacy testing of high-dimensional factor-augmented regression model}
		\end{center}
		\medskip
	} \fi
	
	\bigskip
	\begin{abstract}
	In this paper, we investigate the adequacy testing problem of high-dimensional factor-augmented regression model. Existing test procedures perform not well under dense alternatives. To address this critical issue, we introduce a novel quadratic-type test statistic which can efficiently detect dense alternative hypotheses. We further propose an adaptive test procedure to remain powerful under both sparse and dense alternative hypotheses. 
	Theoretically, under the null hypothesis, we establish the asymptotic normality of the proposed quadratic-type test statistic and asymptotic independence of the newly introduced quadratic-type test statistic and a maximum-type test statistic. We also prove that our adaptive test procedure is powerful to detect signals under either sparse or dense alternative hypotheses. Simulation studies and an application to an FRED-MD macroeconomics dataset are carried out to illustrate the merits of our introduced procedures.	 
	\end{abstract}
	
	\noindent%
	
	\newpage
	\spacingset{1.9} % DON'T change the spacing!

	\section{Introduction}
	\label{sec:intro}

	In this paper, we consider the following Factor Augmented linear Regression Model (FARM, \cite{fanJ2023}):
	\begin{align} 
		Y&=\boldsymbol{f}^{\top}\boldsymbol{\gamma}^*+\boldsymbol{u}^{\top}\boldsymbol{\beta}^*+\varepsilon, \label{linear regression models with hidden confounders} \\
		\text {with} \ \ 
		\boldsymbol{x}&=\boldsymbol{B}\boldsymbol{f}+\boldsymbol{u}.\label{Factormodel}
	\end{align}
	Here, $Y\in \mathbb{R}$  is a response variable, $\boldsymbol{x}$ is a $p$-dimensional covariate vector, $\boldsymbol{f}$ is a $K$-dimensional vector of latent factors, $\boldsymbol{B} \in \mathbb{R}^{p \times K}$ is the corresponding factor loading matrix, and $\boldsymbol{u}$ is a $p$-dimensional vector  of idiosyncratic component, which is uncorrelated with $\boldsymbol{f}$. The $\boldsymbol{\beta}^*=(\beta_1^*,\ldots,\beta_p^*)^\top\in \mathbb{R}^{p}$ and $\boldsymbol{\gamma}^*=(\gamma_1^*,\ldots,\gamma_K^*)^\top\in \mathbb{R}^{K}$ are vectors of regression parameters quantifying the contribution of $\boldsymbol{u}$
	and  $\boldsymbol{f}$, respectively.  The random error $\varepsilon$  satisfies that $\mathbb{E}\left(\varepsilon\right)=0$, and is independent of $\boldsymbol{u}$ and $\boldsymbol{f}$.
	%In fact, model \eqref{Factormodel} is a commonly used structure for characterizing the  interdependence among features. In this framework, the variables are intercorrelated through a shared set of latent factors. %This paper focuses on scenarios where $p$ is high-dimensional while $K$ is fixed.   
    As discussed by \cite{fanJ2023}, the factor augmented regression model encompasses the factor regression model \citep{bai2008forecasting,fan2017sufficient, bing2021prediction} by adding an augmented part $\boldsymbol{u}^{\top}\boldsymbol{\beta}^*$. The above factor augmented regression model has been investigated by many authors \citep{guo2022doubly, ouyang2023high, sun2023decorrelating}. Its high popularity is partly due to the fact that it is a powerful tool to deal with interdependence among covariates.

	%	Numerous methodologies have been proposed to enable  statistical inferences regarding regression parameters in model \eqref{linear regression models with hidden confounders}.
	%		\cite{guo2022doubly} introduced a deconfounding approach for conducting statistical inference on individual regression coefficient ${\beta}_{j}^*, j=1,2,\ldots,p$, integrating the findings of \cite{cevid2020spectral} with the debiased Lasso method. \cite{ouyang2023high}
	%		extended this work to the Generalized Linear Models (GLM). \cite{sun2023decorrelating} addressed GLM with hidden confounders while tackling the issue of multiple testing. \cite{bing2023inference}  focused on inferring high-dimensional multivariate response regression models with hidden confounders. \cite{fan2017sufficient} investigated sufficient dimension reduction  techniques considering hidden confounders, further developed by 
	%		\cite{luo2022inverse} and  \cite{jiang2019sufficient} to accommodate high-dimensional covariate. 

For the above high-dimensional factor augmented regression model, an important problem is whether a factor regression model is adequate. In other words, the primary question of interest
is testing the following hypothesis:
	\begin{align}\label{nullhypo}
		H_0:  \boldsymbol{\beta}^*=\boldsymbol{0} \ \ \text{versus} \ \ H_1:\boldsymbol{\beta}^*\neq \boldsymbol{0}.
	\end{align}
	For this critical testing problem, {\cite{fanJ2023} constructed a maximum-type test statistic based on debiased estimator \citep{zhang2014confidence, van2014} of $\boldsymbol{\beta}^*$. However, their procedure may be computationally expensive since it requires to estimate high-dimensional precision matrix. To reduce the computation burden, \cite{beyhum2023tuning} recently proposed a new maximum-type test statistic without estimating high-dimensional precision matrix. As noted by \cite{beyhum2023tuning}, these maximum-type test statistics exhibit very low power against dense alternatives. This is understandable since the maximum-type test statistics only consider the maximum signal which can be very small under dense alternatives. To achieve high detection power under dense alternatives, we introduce a novel quadratic-type test statistic which integrates the signals. Similar quadratic-type tests also appear in global testing of high-dimensional regression coefficients \citep{zhong2011tests, cui2018test, yang2024score}. However, different from existing test procedures, here the factors ${\boldsymbol{f}}$ and ${\boldsymbol{u}}$ are unobserved, which brings theoretical challenges.

In practice, we actually do not know whether the alternatives are sparse or dense. When the parametric vector $\boldsymbol{\beta}^*$ is extremely sparse, the power of quadratic-type test statistics can be low \citep{xu2016adaptive, feng2022max, yu2024fisher, yu2024power}. This further motivates us to develop an adaptive test procedure which works well for both sparse and dense alternatives. After establishing the asymptotic normality of the proposed quadratic-type test statistic, we prove that the quadratic-type test and the maximum-type test in \cite{beyhum2023tuning} are asymptotically independent. This important result enables us to propose an adaptive test procedure to combine the strength of both quadratic-type test and maximum-type test. We demonstrate that the adaptive test procedure is powerful to detect signals under either sparse or dense alternative hypotheses. We also make some modifications of the test statistic in \cite{beyhum2023tuning}. We derive the asymptotic distribution of our modified statistic. Our modification avoids the estimation of high dimensional regression coefficient $\boldsymbol{\beta}^*$, which is required for the procedure in \cite{beyhum2023tuning}. 
		
		We make the following important contributions:
		\begin{itemize}
			\item We propose a novel quadratic-type test statistic which is powerful for dense alternatives and we establish its asymptotic normality;
			\item We establish the asymptotic independence of the introduced quadratic-type test and the maximum-type
			test in \cite{beyhum2023tuning};
			\item We propose an adaptive test procedure and demonstrate that the adaptive test procedure is powerful to detect signals under either sparse or dense alternative hypotheses. 
		\end{itemize}
Although the quadratic-type tests have been introduced and investigated by many authors, to our best knowledge, quadratic-type tests with estimated latent factors have not been investigated. The theoretical investigation in this paper is very challenging. Actually since the factors are unobserved, we have to carefully deal with the accumulated estimation errors of the estimated latent factors. Our theoretical investigation in this paper would be very helpful for other relevant researches.

The rest of the paper is organized as follows. In Section \ref{Methodology}, we construct novel test statistics. In Section \ref{Theoretical results section},  we present asymptotic properties of our constructed test statistics.
		We present our simulation studies in Section \ref{simulations} and provide an analysis of real data in Section \ref{Real data analysis section} to assess the performance of the proposed approaches. Conclusions and discussions are presented in Section \ref{Conclusions and discussions section}. All technical proofs are given in the Supplementary Material.

		\textbf{Notation.} Let $\mathbb{I}(\cdot)$  denote the indicator function. For a vector $\boldsymbol{a}=\left(a_{1},\ldots, a_{m}\right)^{\top}\in \mathbb{R}^{m}$, we denote its $\ell_{q}$ norm as $\|\boldsymbol{a}\|_{q}=\left(\sum_{\ell=1}^{m}|a_{\ell}|^{q}\right)^{1/q}, \ 1\leq q< \infty$, $\|\boldsymbol{a}\|_{\infty}=\max_{1\leq \ell \leq m}|a_{\ell}|$,  and $\|\boldsymbol{a}\|_{0}=\sum_{\ell=1}^{m}\mathbb{I}(a_{\ell}\neq 0)$.
		For any integer $m$, we define $[m]=\{1,\ldots, m\}$.
		The sub-Gaussian norm  of a scalar random variable $X$ is defined as $\|X\|_{\psi_{2}}=\inf{\{c>0: \mathbb{E}\exp(X^{2}/c^{2})\leq 2\}}$.
        The sub-Exponential norm of a random variable $X$ is defined as $\|X\|_{\psi_1}=\inf\{t>0: \mathbb{E}\exp(|X|/t) \leq 2\}$.
		For a random vector $\boldsymbol{x}\in \mathbb{R}^{m}$, we define its  sub-Gaussian norm  as $\|\boldsymbol{x}\|_{\psi_{2}}=\sup_{\|\boldsymbol{c}\|_{2}=1}\|\boldsymbol{c}^{\top}\boldsymbol{x}\|_{\psi_{2}}$.
		Furthermore, we use $\boldsymbol{I}_{K}$,  $\boldsymbol{1}_{K}$ and $\boldsymbol{0}_{K}$ to denote the identity matrix in $\mathbb{R}^{K \times K}$, a vector of dimension $K$ with all elements being $1$ and all elements being $0$, respectively.
		For a matrix $\boldsymbol{A}={[A_{jk}]}$, 
		{ $\boldsymbol{a}_{j}$ represents $j$-th row of $\boldsymbol{A}$, and $\boldsymbol{A}_{k}$ represents $k$-th column of $\boldsymbol{A}$.}
		We define $\|\boldsymbol{A}\|_{\mathbb{F}}=\sqrt{\sum_{jk}{A}_{jk}^{2}}$, $\|\boldsymbol{A}\|_{\max}=\max_{j,k}|A_{jk}|$, $\|\boldsymbol{A}\|_{\infty}=\max_{j}\sum_{k}|A_{jk}|$ and $\|\boldsymbol{A}\|_{1}=\max_{k}\sum_{j}|A_{jk}|$
		to be its Frobenius norm, element-wise max-norm, matrix $\ell_{\infty}$-norm and matrix $\ell_{1}$-norm, respectively.
		In addition, we use $\lambda_{\min}(\boldsymbol{A})$ and $\lambda_{\max}(\boldsymbol{A})$  to denote the minimal and maximal eigenvalues of $\boldsymbol{A}$, respectively. We define
        $\tr(\boldsymbol{A})$ as the trace of  $\boldsymbol{A}$.
		We use $|\mathcal{A}|$ to denote the cardinality of a set  $\mathcal{A}$.
		For two positive sequences $\{a_{n}\}_{n \geq 1}$, $\{b_{n}\}_{n \geq 1}$, we write $a_{n}=O(b_{n})$ if there exists a positive constant $C$ such that $|a_{n}/ b_{n}|\leq C$ when $n$ is large enough, and we write $a_{n}=o(b_{n})$ if $a_{n}/b_{n}\rightarrow 0$. 
		Furthermore, if $a_{n}=O(b_{n})$ is satisfied, we write $a_n\lesssim b_n$. If
		$a_n\lesssim b_n$ and $b_n\lesssim a_n$, and we will write it as $a_n\asymp b_n$ for short.
		In addition, let $a_{n}=O_{\mathbb{P}}(b_{n})$ denote $\Pr(|a_n/b_n| \leq c)\rightarrow 1$ for some constant $c < \infty$. Let $a_{n}=o_{\mathbb{P}}(b_{n})$ denote $\Pr(|a_n/b_n| > c)\rightarrow 0$ for any constant $c >0$.
		
		\section{Construction of test statistics} \label{Methodology}
		
		%\subsection{Factor estimation}\label{Factor estimation}
		Throughout the paper, we assume that the data $\{\boldsymbol{x}_{i}, \boldsymbol{f}_{i},  \boldsymbol{u}_{i}, \varepsilon_i,  Y_{i}\}_{i=1}^n$ are  independent and identically distributed (i.i.d.) copies of $\{\boldsymbol{x}, \boldsymbol{f},
		\boldsymbol{u}, {\varepsilon}, 
		Y\}$. However, we should note that only $\{\boldsymbol{x}_{i}, Y_{i}\}_{i=1}^n$ are observable. Let $\boldsymbol{X}=(\boldsymbol{x}_{1}, \ldots, \boldsymbol{x}_{n})^{\top} \in \mathbb{R}^{n \times p}$,  $\boldsymbol{F}=(\boldsymbol{f}_{1}, \ldots, \boldsymbol{f}_{n})^{\top}\in \mathbb{R}^{n \times K}$, $\boldsymbol{U}=(\boldsymbol{u}_{1}, \ldots, \boldsymbol{u}_{n})^{\top}\in \mathbb{R}^{n \times p}$, $\boldsymbol{\varepsilon}=(\varepsilon_{1}, \ldots, \varepsilon_{n})^{\top}\in \mathbb{R}^{n}$ and $\boldsymbol{Y}=(Y_{1}, \ldots, Y_{n})^{\top} \in \mathbb{R}^{n}$.

		%Given that only the predictor vector ${\boldsymbol{x}}$ and $Y$ are observable, the latent factor $\boldsymbol{f}$, its corresponding loading
		%	matrix $\boldsymbol{B}$, and idiosyncratic component $\boldsymbol{u}$ are not identifiable under factor model \eqref{Factormodel}. Therefore, it is necessary to estimate the factor model. 
		To make inference on the regression coefficient $\boldsymbol{\beta}^*$, we need to first estimate the latent factors $\boldsymbol{f}$ and $\boldsymbol{u}$. Firstly, let $\hat{K}$ be one of the many consistent estimators of the number of factors $K$ available in the literature \citep{BaiandNg2002, Lam2012, Ahn2013, Fan2022}.
		As in \cite{stock2002forecasting, Bai2003, fanJ2023}, let  the columns of $\hat{\boldsymbol{F}}/\sqrt{n}$ be the eigenvectors corresponding to the largest $\hat{K}$ eigenvalues of the matrix ${\bX}{{\bX}}^{\top}$ and $\hat{\boldsymbol{B}}={\bX}^{\top}{\hat{\boldsymbol{F}}}({\hat{\boldsymbol{F}}}^{\top}\hat{\boldsymbol{F}})^{-1}=n^{-1}{\bX}^{\top}{\hat{\boldsymbol{F}}}$.
		Then the estimator of $\boldsymbol{U}$ is  $$\hat{\boldsymbol{U}}={\bX}-\hat{\boldsymbol{F}}{\hat{\boldsymbol{B}}}^{\top}=\left(\boldsymbol{I}_{n}-\frac{1}{n}\hat{\boldsymbol{F}}{\hat{\boldsymbol{F}}}^{\top}\right){\bX}=(\boldsymbol{I}_{n}-\hat{\boldsymbol{P}}){\bX}.$$
		Here, $\hat{\boldsymbol{P}}=n^{-1}\hat{\boldsymbol{F}}\hat{\boldsymbol{F}}^{\top}$ is the corresponding projection matrix.

		Under the null hypothesis in (\ref{nullhypo}), we have 
		$\mathbb{E}\{\boldsymbol{u}(Y-\boldsymbol{f}^{\top}\boldsymbol{\gamma}^*)\}=\mathbb{E}(\boldsymbol{u}\varepsilon)=\boldsymbol{0}$. While under the alternative hypothesis,  we have $\mathbb{E}\{\boldsymbol{u}(Y-\boldsymbol{f}^{\top}\boldsymbol{\gamma}^*)\}=\mathbb{E}\{\boldsymbol{u}(\boldsymbol{u}^{\top}\boldsymbol{\beta}^*+\varepsilon)\}=\boldsymbol{\Sigma}_{\boldsymbol{u}}\boldsymbol{\beta}^* \neq \boldsymbol{0}$. Here $\boldsymbol{\Sigma}_{\boldsymbol{u}}=\mathbb{E}(\boldsymbol{u}\boldsymbol{u}^{\top})$. This motivates us to construct test statistics based on empirical versions of norms of  $\mathbb{E}\{\boldsymbol{u}(Y-\boldsymbol{f}^{\top}\boldsymbol{\gamma}^*)\}$.
		
		Based on the above observation, \cite{beyhum2023tuning} investigated the following test statistic
		$$S_{n}=\max_{1\leq j\leq p}\left|\frac{1}{\sqrt n}\sum_{i=1}^n\widehat U_{ij} (Y_i-\hat{\boldsymbol{f}}_i^{\top}\hat{\boldsymbol{\gamma}})\right|.$$
		Here, $\hat{\boldsymbol{\gamma}}=(\hat{\boldsymbol{F}}^{\top}\hat{\boldsymbol{F}})^{-1}\hat{\boldsymbol{F}}^{\top}{\boldsymbol{Y}}=n^{-1}\hat{\boldsymbol{F}}^{\top}{\boldsymbol{Y}}$,  $\hat{\boldsymbol{f}}_i$ and $\hat{{U}}_{ij}$ denote the $i$-th row and the $(i,j)$-th element of the estimated matrices $\hat{\boldsymbol{F}}$ and $\hat{\boldsymbol{U}}$, respectively.  In \cite{beyhum2023tuning}, to determine the critical value of $S_n$, they resorted to Gaussian multiplier bootstrap and required to estimate high-dimensional parametric vector $\boldsymbol{\beta}^*$. We argue that for the above $S_n$, we actually do not need to estimate $\boldsymbol{\beta}^*$. To this end, 
        we make some modification for the above $S_n$. Now denote $\hat{\sigma}_j^2=n^{-1}\sum_{i=1}^n\hat{U}_{ij}^2(Y_i-\hat{\boldsymbol{f}}_i^{\top}\hat{\boldsymbol{\gamma}})^2$.	 We show that $\hat{\sigma}_j^2 $ is a   consistent estimator of $\sigma_j^{\natural2}=\mathbb{E}\{U_{ij}^2(Y_i-{\boldsymbol{f}}_{i}^{\top}{\boldsymbol{\gamma}}^*)^2\}$ by Lemma \ref{total variance estimation error bound lemma} in  Supplementary Material.	%$\hat{\boldsymbol{\Sigma}}_{u}=n^{-1}\sum_{i=1}^{n}\hat{\boldsymbol{u}}_{i}\hat{\boldsymbol{u}}_{i}^{\top}$ and  $\widehat\sigma_j^2=\hat{\sigma}_{\varepsilon}^2\hat{\boldsymbol{\Sigma}}_{u, jj}$ with  
		%$\hat{\sigma}_{\varepsilon}^2=n^{-1}\sum_{i=1}^n(Y_i-\hat{\boldsymbol{f}}_{i}^{\top}\hat{\boldsymbol{\gamma}})^2$. 
        %By Lemma \ref{total variance estimation error bound lemma}, we know that $\widehat\sigma_{j}^2$ is a consistent estimator of 
		Then we define $$\tilde{S}_n=\max_{1\leq j\leq p}\left|\frac{1}{\sqrt{n\widehat\sigma_{j}}}\sum_{i=1}^{n}\widehat U_{ij} (Y_i-\hat{\boldsymbol{f}}_i^{\top}\hat{\boldsymbol{\gamma}})\right|.$$  
		Clearly the above $\tilde{S}_n$ is a standardized version of the $S_n$. For the above modified test statistic $\tilde S_n$, we can derive its asymptotic distribution. Actually Theorem \ref{max statistic theorem} in Section \ref{Theoretical results section} shows that the $\tilde{S}_n^2$ converges to the Gumbel distribution, with the cumulative distribution function    $F(y)=\exp\{-\pi^{-1/2}\exp(-y/2)\}$. We then reject the $H_0$ at significance level $\alpha$ if $\tilde{S}_n^2  
        \geq c(\alpha)$, where $c(\alpha)=2 (\log p)-\{\log( \log p)\}+q_{\alpha}$ and $q_{\alpha}$  is the $(1-\alpha)$-quantile of the Gumbel distribution, that is, $q_\alpha=-(\log \pi)-2[\log \{\log (1-\alpha)^{-1}\}]$.
		
		%Further recall that $\boldsymbol{u}$ and $\boldsymbol{f}$ are uncorrelated. Then it follows that under the null hypothesis, we have $\mathbb{E}[\boldsymbol{u}^{\top}Y]=0$. 
		
		%For the parametric vector $\boldsymbol{\gamma}$, we estimate it as follows
		%\begin{align}
		%			\hat{\boldsymbol{\gamma}}&=(\hat{\boldsymbol{F}}^{\top}\hat{\boldsymbol{F}})^{-1}\hat{\boldsymbol{F}}^{\top}\boldsymbol{Y}. \label{the least squares estimate of phi}
		%		\end{align}
	
	As noted by \cite{beyhum2023tuning}, the above maximum-type test statistic $S_n$ exhibits low power against dense alternatives. To address this critical issue, we introduce a novel  quadratic-type test statistic. Actually now we consider the following test statistic
	$$T_{n}=\frac{1}{n(n-1)}\sum_{i\neq j}^n(Y_i-\hat{\boldsymbol{f}}_i^{\top}\hat{\boldsymbol{\gamma}})(Y_j-\hat{\boldsymbol{f}}_j^{\top}\hat{\boldsymbol{\gamma}})\hat{\boldsymbol{u}}_i^{\top}\hat{\boldsymbol{u}}_j.$$
Here, $\hat{\boldsymbol{u}}_i$ is the $i$-th row of $\hat{\boldsymbol{U}}$. Clearly the above $T_n$ is an empirical version of the $\ell_2$ norm of $\mathbb{E}\{\boldsymbol{u}(Y-\boldsymbol{f}^{\top}\boldsymbol{\gamma}^*)\}$. While $S_n$ considers the $\ell_{\infty}$ norm of $\mathbb{E}\{\boldsymbol{u}(Y-\boldsymbol{f}^{\top}\boldsymbol{\gamma}^*)\}$. For dense alternative hypotheses, the $\ell_{\infty}$ norm of $\mathbb{E}\{\boldsymbol{u}(Y-\boldsymbol{f}^{\top}\boldsymbol{\gamma}^*)\}$ can be small, but the $\ell_2$ norm of $\mathbb{E}\{\boldsymbol{u}(Y-\boldsymbol{f}^{\top}\boldsymbol{\gamma}^*)\}$ can be large. This makes the above test statistic $T_{n}$  powerful for dense alternative hypotheses.
	
	To formulate a test procedure based on $T_n$, we need to estimate $\mbox{tr}(\boldsymbol{\Sigma}^2_{\boldsymbol{u}})$ appeared in the asymptotic variance. 
	%We will use the estimator of $\mbox{tr}(\boldsymbol{\Sigma}^2_{\boldsymbol{u}})$ proposed in \cite{chen2010tests}.
	%Denoted $P_n^m=n!/(n-m)!$, a ratio consistent estimator of $\mbox{tr}(\boldsymbol{\Sigma}^2_{\boldsymbol{u}})$ is 
	Proposition \ref{tracetilde consistent property} shows that 
	$\hat{\mbox{tr}(\boldsymbol{\Sigma}^2_{\boldsymbol{u}})}=\{n(n-1)\}^{-1}\sum_{i \neq j}(\hat{\boldsymbol{u}}_{i}^{\top}\hat{\boldsymbol{u}}_{j})^2$ is a ratio consistent estimator of $\mbox{tr}(\boldsymbol{\Sigma}^2_{\boldsymbol{u}})$.
	%-\frac{2}{P_n^3}\sum\limits_{i_1 \neq i_2 \neq i_3}\hat{\boldsymbol{u}}_{i_1}^{\top}\hat{\boldsymbol{u}}_{i_2}\hat{\boldsymbol{u}}_{i_2}^{\top}\hat{\boldsymbol{u}}_{i_3}+\frac{1}{P_n^4}\sum\limits_{i_1 \neq i_2 \neq i_3 \neq i_4}\hat{\boldsymbol{u}}_{i_1}^{\top}\hat{\boldsymbol{u}}_{i_2}\hat{\boldsymbol{u}}_{i_3}^{\top}\hat{\boldsymbol{u}}_{i_4}.$$
	%The estimator  of $\sigma_{\varepsilon}^2$ under $H_0$ is 
	%\begin{align}
	%\hat{\sigma}_{\varepsilon}^2=\frac{1}{n}\sum_{i=1}^n\hat{\varepsilon}_i^2
	%\end{align}
	Denote $M_n=nT_n/\sqrt{{2\hat{\mbox{tr}(\boldsymbol{\Sigma}^2_{\boldsymbol{u}})}}\hat{\sigma}_{\varepsilon}^4}$. Here $\hat{\sigma}_{\varepsilon}^2=n^{-1}\sum_{i=1}^n(Y_i-\hat{\boldsymbol{f}}_i^{\top}\hat{\boldsymbol{\gamma}})^2$.
	Applying Theorem \ref{sum statistic normal distribution} and Slutsky's Theorem,  the proposed test rejects $H_0$ at a significant level $\alpha$ if $M_n\geq z(\alpha)$, where $z(\alpha)$ is the upper $\alpha$-quantile of standard Gaussian distribution $\mathrm{N}(0,1)$.

	\iffalse  
	\begin{figure}[H]
		\centering
		\includegraphics[width=0.4\textwidth]{hat_eiejuiuj.eps}
		\includegraphics[width=0.4\textwidth]{eiejuiuj.eps}
		\caption{Left: $\frac{1}{n(n-1)}\sum_{i\neq j}\hat{\varepsilon}_i\hat{\varepsilon}_j\hat{\boldsymbol{u}}_i^\top \hat{\boldsymbol{u}}_j$; Right: $\frac{1}{n(n-1)}\sum_{i\neq j} {\varepsilon}_i {\varepsilon}_j {\boldsymbol{u}}_i^\top  {\boldsymbol{u}}_j$. (after  scaling)
		} 
	\end{figure} \fi
	\iffalse
	\begin{figure}
		\centering
		\includegraphics[width=0.4\linewidth]{Tn_YiYj.eps}
		\includegraphics[width=0.4\linewidth]{A1.eps}\caption{Left: $\frac{1}{n(n-1)}\sum_{i\neq j}Y_iY_j\hat{\boldsymbol{u}}_i^\top \hat{\boldsymbol{u}}_j$; Right: $\frac{1}{n(n-1)}\sum_{i\neq j} Y_i Y_j {\boldsymbol{u}}_i^\top  {\boldsymbol{u}}_j$. (after  scaling)}
		\label{fig:enter-label}
	\end{figure} \fi

	%For both $T_n$ and $S_n$, we do not need to estimate the high-dimensional parametric vector $\boldsymbol{\beta}^*$. %We also do not need to estimate the low-dimensional parametric vector $\boldsymbol{\gamma}^*$. Thus our procedure can be computationally efficient. 
	%Since $\boldsymbol{\gamma}^*$ is of size $K$, it is reasonable to assume that $\|\boldsymbol{\gamma}^*\|_2=O(1)$.
	%Let  $\mathcal{S}^*=\left\{j \in [p]: \beta_{j}^* \neq 0\right\}$ and $s=|\mathcal{S}^*|$ be its cardinality.
	
As discussed above, the maximum-type test statistic $\tilde S_n^2$ and the quadratic-type test statistic $M_n$ are powerful against sparse and dense alternative hypotheses, respectively. However, in practice we do not know the sparsity level of alternative hypotheses. This fact then asks us to develop an adaptive test procedure which could be powerful under both sparse and dense alternative hypotheses. 
To achieve this goal, the following adaptive test procedure is introduced. Our adaptive test procedure is actually the Fisher adaptive test statistic, denoted by  $F_n$, which combines the two $p$-values as 
	\begin{align} \label{pvalue combine}
		F_n=-2\{ (\log p_{M})+(\log p_{S})\}.
	\end{align}
	Here, $p_M=1-\Phi(M_n)$ and $p_S=1-F[\tilde{S}_n^2-2(\log p)+\{\log( \log p)\}]$, with $\Phi(\cdot)$ and $F(\cdot)$ being the  cumulative distribution functions  of standard normal distribution and Gumbel distribution, respectively. 
    %and $F(y)=\exp\{-\pi^{-1/2}\exp(-y/2)\}$ being the cdf of a Gumbel distribution.
The above adaptive test statistic is motived by the result in Theorem \ref{sum and max independent theorem}. Actually Theorem \ref{sum and max independent theorem} in Section \ref{Theoretical results section} establishes the asymptotic independence of the maximum-type statistic $\tilde{S}_n^2$ and the quadratic-type test statistic $M_n$. This important result enables us to combine the merits of both $\tilde{S}_n^2$ and $M_n$. Given the asymptotic independence in  Theorem \ref{sum and max independent theorem}, under the null hypothesis $H_0$, $F_n$  converges to  a chi-square distribution with four degrees of freedom, $\chi_4^2$,  as $n,p \rightarrow \infty$. The adaptive test procedure rejects $H_0$ if $F_n \geq \chi_{\alpha}$, where $\chi_{\alpha}$ is the upper $\alpha$-quantile of the
	$\chi_4^2$ distribution. 
    %Fisher adaptive test rejects the  null hypothesis if and only if $\Psi_{\alpha}^{F}=1$.
    We will formally prove that the adaptive test procedure is powerful to detect signals under either sparse or dense alternative hypotheses.

	\section{Theoretical results}     \label{Theoretical results section}  
	In this section, we derive the asymptotic distributions of the quadratic-type test statistic $T_n$ and the maximum-type test statistic $\widetilde S_n^2$ and establish the asymptotic independence of $T_n$ and $\widetilde S_n^2$. To begin, we impose the identifiability and regularity assumptions commonly adopted in the literature on factor analysis, including those in \cite{Bai2003, FanJ2013, fanJ2023}.
	%, and other notable literature concerning
	%high-dimensional factor analysis.
	%\begin{assumption} \label{basic asssumption in factor model}
	%Assume that $\Cov(\boldsymbol{f})=\boldsymbol{I}_{K}$ and $\boldsymbol{B}^{\top}\boldsymbol{B}$ is diagonal.
	%\end{assumption}

	\begin{assumption} \label{factorassumption} 
		It holds that 		
		\begin{itemize}
			
			\item[(i)] $\mathbb{E}(\boldsymbol{f})=\boldsymbol{0}_{K}$, $\mathbb{E}(\boldsymbol{u})=\boldsymbol{0}_{p}$, $\Cov(\boldsymbol{f})=\boldsymbol{I}_{K}$ and $\boldsymbol{B}^{\top}\boldsymbol{B}$ is diagonal.
			\item[(ii)] $\|\boldsymbol{f}\|_{\psi_{2}}\leq c_{0}$, $\|\boldsymbol{u}\|_{\psi_{2}}\leq c_{0}$ and $\|{\boldsymbol{u}}^{\top}{\boldsymbol{\beta}}^*\|_{\psi_2} \leq c_0$ for some positive constant $c_0$. 
			\item[(iii)] The eigenvalues of  $p^{-1}\boldsymbol{B}^{\top}\boldsymbol{B} $  remain uniformly bounded away from both zero and infinity as $p \rightarrow \infty$.
			\item[(iv)]  $\|\boldsymbol{B}\|_{\max}=O(1)$ and $\|\boldsymbol{\Sigma}_{\boldsymbol{u}}\|_{1}=O(1)$.
			%and $$\mathbb{E}|\boldsymbol{u}^{\top}\boldsymbol{u}-\tr(\boldsymbol{\Sigma}_{\boldsymbol{u}})|^{4}\leq \Upsilon p^{2}.$$ 
			\item[(v)] 
			% There exist a positive constant $\kappa<1$ such that $\kappa \leq \lambda_{\min}(\boldsymbol{\Sigma}_{\boldsymbol{u}}), \ \lambda_{\max}(\boldsymbol{\Sigma}_{\boldsymbol{u}})$,
			There exist  positive  constants $c_1< c_2$ such that $c_1< \lambda_{\min}(\boldsymbol{\Sigma}_{\boldsymbol{u}}) \leq  \lambda_{\max}(\boldsymbol{\Sigma}_{\boldsymbol{u}}) \leq c_2$ and   $\min_{1\leq k,l\leq p} \Var({U}_{ik}{U}_{il})> c_1$.
			
		\end{itemize}
	\end{assumption}
	
	In addition, let $\hat{\boldsymbol{\Sigma}}$, $\hat{\boldsymbol{\Lambda}}=\text{diag}(\hat{\lambda}_{1},\ldots, \hat{\lambda}_{K})$ and $\hat{\boldsymbol{\Gamma}}=(\hat{\boldsymbol{\zeta}}_{1},\ldots, \hat{\boldsymbol{\zeta}}_{K})$ be  estimators of the covariance matrix $\boldsymbol{\Sigma}$ of $\boldsymbol{x}$,  the matrix consisting of its leading $K$ eigenvalues $\boldsymbol{\Lambda}=\text{diag}(\lambda_{1},\ldots, \lambda_{K})$, and the matrix consisting of their corresponding $K$  orthonormalized eigenvectors $\boldsymbol{\Gamma}=(\boldsymbol{\zeta}_{1},\ldots, \boldsymbol{\zeta}_{K})$, respectively. 
	%Following \cite{fan2018large}, we take 
	%$\{\hat{\lambda}_{j}\}_{j=1}^K$ and $\{\hat{\boldsymbol{\zeta}}_{j}\}_{j=1}^K$ 
	% in modulus form to enhance adaptability across different applications.  In other words, they are not necessarily the eigenvalues and corresponding eigenvectors of $\hat{\boldsymbol{\Sigma}}$. 
	For $\hat{\boldsymbol{\Sigma}}$, $\hat{\boldsymbol{\Lambda}}$, $\hat{\boldsymbol{\Gamma}}$, the following assumption is imposed.

	\begin{assumption} \label{Loadings and initial pilot estimators}
		\iffalse Assume that 
		$\hat{\boldsymbol{\Sigma}}$, $\hat{\boldsymbol{\Lambda}}$ and $\hat{\boldsymbol{\Gamma}}$ satisfy $\|\hat{\boldsymbol{\Sigma}}-\boldsymbol{\Sigma}\|_{\max}=O_{\mathbb{P}}\{\sqrt{(\log p)/n}\}$, $\|(\hat{\boldsymbol{\Lambda}}-\boldsymbol{\Lambda})\hat{\boldsymbol{\Lambda}}^{-1}\|_{\max}=O_{\mathbb{P}}\{\sqrt{(\log p)/n}\}$, and $\|\hat{\boldsymbol{\Gamma}}-\boldsymbol{\Gamma}\|_{\max}=O_{\mathbb{P}}\{\sqrt{(\log p)/(np)}\}$. \fi
		\begin{align*}
			\|\hat{\boldsymbol{\Sigma}}-\boldsymbol{\Sigma}\|_{\max}&=O_{\mathbb{P}}\left\{\sqrt{(\log p)/n}\right\}, \notag \\
			\|(\hat{\boldsymbol{\Lambda}}-\boldsymbol{\Lambda})\hat{\boldsymbol{\Lambda}}^{-1}\|_{\max}&=O_{\mathbb{P}}\left\{\sqrt{(\log p)/n}\right\}, \notag \\
			\|\hat{\boldsymbol{\Gamma}}-\boldsymbol{\Gamma}\|_{\max}&=O_{\mathbb{P}}\left\{\sqrt{(\log p)/(np)}\right\}. \notag 
		\end{align*}
	\end{assumption}
	The above assumption holds in various relevant settings. Actually, it applies to the sample covariance matrix under sub-Gaussian distributions \citep{FanJ2013}, as well as to estimators such as the marginal and spatial Kendall's tau \citep{fan2018large}.
	We summarize the theoretical results related to consistent factor estimation in Lemma \ref{consistent factor estimation} in Supplementary Material,  which directly follows from Proposition 2.1 in \cite{fanJ2023} and Lemma 3.1 in \cite{bayle2022factor}.
To investigate the asymptotic performance, we impose the	following sub-Gaussian condition on the random noise $\varepsilon$, which is commonly adopted in the literature \citep{fanJ2023, yang2024score,  shi2025testing}.

	\begin{assumption}\label{error sub-Gaussian assumption}
		There exists a positive constant $c_{1}$ such that $\|\varepsilon\|_{\psi_2} \leq c_1$. 
	\end{assumption}
	%\begin{assumption}\label{Sigmanatural bound assumption}
	%Assume that $\lambda_{\max}(\boldsymbol{\Sigma}_{\boldsymbol{u}})/\sqrt{\text{tr}(\boldsymbol{\Sigma}_{\boldsymbol{u}}^{2})}=o\{(\log p)^{-1-\alpha_0}\}$ for a constant $\alpha_0 >0$.
	%\end{assumption}
	%Assumption \ref{error sub-Gaussian assumption}  states that $\varepsilon$ is a sub-Gaussian variable. This is a mild condition 
	%The condition $\lambda_{\max}(\boldsymbol{\Sigma}_{\boldsymbol{u}})/\sqrt{\text{tr}(\boldsymbol{\Sigma}_{\boldsymbol{u}}^{2})}\rightarrow 0$ is sufficient for establishing the
	%central limit theorem for quadratic-type test statistics $T_n$, which was used in \cite{chen2010two, li2024power}.
	
Denote $\sigma^2_{\varepsilon}=\mathbb{E}\{(Y-f^\top\boldsymbol{\gamma}^*)^2\}$.  Then with the above assumptions, we have the following theorem.
	\begin{theorem} \label{sum statistic normal distribution}
		Suppose  that $p\asymp n^{a},\ 2/3<a <2$, and $\|\boldsymbol{B}_{k}^{\top}{\boldsymbol{x}}\|_{\psi_2}\leq c$ for $k \in [K]$ and some constant $c$. Under Assumptions \ref{factorassumption}-\ref{error sub-Gaussian assumption} and the null hypothesis $H_0$,  %if $(\log p)(\log n)=o(\sqrt{n})$ and $\log n =o(\sqrt{p})$, 
		%if $s\left[{(\log p)^2\sqrt{ \log n}}/{p}+\{(\log p)^2(\log n)^{3/2}\}/{\sqrt{n}}\right]=O(1)$,
		we have 
		\begin{align}
			\frac{{n}T_n}{\sigma_{\varepsilon}^2\sqrt{2\mbox{tr}(\boldsymbol{\Sigma}^2_{\boldsymbol{u}})}} \rightarrow \mathrm{N}(0,1). \notag 
		\end{align}
		%Here  $\boldsymbol{\Sigma}_u=\mathbb{E}(\boldsymbol{u}\boldsymbol{u}^\top)$.
	\end{theorem}
	
	\iffalse 
	\begin{assumption}\label{moment condition}
		Let $\boldsymbol{A}^{(1)}$ and $\boldsymbol{A}^{(2)}$ be two $p \times p$ positive semi-definite matrices. Assume that 
		$\mathbb{E}\{\Pi_{i=1}^2 \boldsymbol{u}^{\top}\boldsymbol{A}^{(i)}\boldsymbol{u}\} \leq C\Pi_{i=1}^2 \tr\{\boldsymbol{A}^{(i)}\boldsymbol{\Sigma}_{\boldsymbol{u}}\}$.
	\end{assumption} \fi

	In the above theorem, we establish the limiting null distribution of the proposed quadratic-type test statistic $T_n$. To obtain the asymptotic normality, we make some restrictions on the dimension $p$, which is inevitable. Actually in the proof, we decompose  $T_n=T_n^*+\triangle_n$, with $T_n^*=\{n(n-1)\}^{-1}\sum_{i\neq j}\varepsilon_i\varepsilon_j{\boldsymbol{u}}_{i}^{\top}{\boldsymbol{u}}_{j}$ and $\triangle_n=T_n-T_n^*=O_{\mathbb{P}}\{{p(\log p)^2(\log n)^4}/{n^2}+{(\log p)^{5/2}}/{\sqrt{np}}+$  ${(\log n)}/{p}\}$.
	On the one hand, 
	the order ${p(\log p)^2(\log n)^4}/{n^2} $ of 
	the estimation errors from terms containing $(\hat{\boldsymbol{u}}_i-{\boldsymbol{u}}_i)^\top(\hat{\boldsymbol{u}}_i-{\boldsymbol{u}}_i)$ must be controlled. Although we have $\max _{i \in [n]}\left\|\hat{\boldsymbol{u}}_{i}-\boldsymbol{u}_{i}\right\|_{\infty}=O_{\mathbb{P}}\{\sqrt{{(\log n)}/{p}}+\sqrt{{(\log p)(\log n)}/{n}}\}$ under mild conditions, this term $(\hat{\boldsymbol{u}}_i-{\boldsymbol{u}}_i)^\top(\hat{\boldsymbol{u}}_i-{\boldsymbol{u}}_i)$ still can be large due to the accumulation over $p$ dimensions. On the other hand, 
    in factor analysis, it is typically assumed that $\sqrt{n} =o(p)$ \citep{Bai2003, bai2006confidence, corradi2014testing, beyhum2023tuning}. To derive the asymptotic normality of quadratic-type test statistic, it is also common to require that $p\rightarrow \infty$ \citep{cui2018test, yang2024score}. Both $\max_{i \in [n]}\|\hat{\boldsymbol{f}}_i-{\boldsymbol{f}}_i\|_2$ and $\|\hat{\boldsymbol{B}}-{\boldsymbol{B}}\|_{\max}$ are of the order $\sqrt{(\log n)/p}+\sqrt{(\log p)(\log n)/n}$.  In the decomposition of the quadratic-type test statistic, the combination of the estimation errors from terms involving $\max_{i \in [n]}\|\hat{\boldsymbol{f}}_i-{\boldsymbol{f}}_i\|_2$ and $\|\hat{\boldsymbol{B}}-{\boldsymbol{B}}\|_{\max}$ results in the order of ${(\log p)^{5/2}}/{\sqrt{np}}+{(\log n)}/{p}$. 
	To ensure that this estimation  error is asymptotically negligible compared to the leading term $T_n^*$, the condition $n^{2/3}=o(p)$  is required. Moreover, it is worth noting that in their Assumption 2.2, \cite{fanJ2023} imposed the dimensional constraint  $n(\log n)^2=O(p)$. Our assumption encompasses this setting  but also allows for a broader range of dimensions. %While in \cite{beyhum2023tuning}, they required that $\sqrt{n} =o(p)$. 

	For inference, we need to estimate $\mbox{tr}(\boldsymbol{\Sigma}^2_{\boldsymbol{u}})$. The following 
	Proposition \ref{tracetilde consistent property} establishes the ratio consistency of 
	$\hat{\mbox{tr}(\boldsymbol{\Sigma}^2_{\boldsymbol{u}})}=\{n(n-1)\}^{-1}\sum_{i \neq j}(\hat{\boldsymbol{u}}_{i}^{\top}\hat{\boldsymbol{u}}_{j})^2$ to $\mbox{tr}(\boldsymbol{\Sigma}^2_{\boldsymbol{u}})$. 
	\begin{proposition} \label{tracetilde consistent property}
		Suppose that the conditions in Theorem \ref{sum statistic normal distribution} hold, then 
		\begin{align}
			\frac{\hat{\mbox{tr}(\boldsymbol{\Sigma}^2_{\boldsymbol{u}})}}{\mbox{tr}(\boldsymbol{\Sigma}_{\boldsymbol{u}}^2)} \rightarrow 1, \ \text{as} \ \  (n,p) \rightarrow \infty. \notag 
		\end{align}
	\end{proposition}
	Theorem \ref{sum statistic normal distribution} and Proposition \ref{tracetilde consistent property} imply that our quadratic-type test statistic $M_n$ can control the size asymptotically. Now we come to investigate the power performance of the quadratic-type test statistic $T_n$. We need to assume the following condition. 
	\begin{assumption} \label{Power1 assumption}
		Assume that $\|\boldsymbol{\beta}^*\|_1(1/{\sqrt{n}}+{p}/{n^{3/2}}+{1}/{p})=o(\|\boldsymbol{\beta}^*\|_2^2)$.
		%$=o(\|\boldsymbol{\beta}^*\|_2^2)$.
		%$\|\boldsymbol{\beta}^*\|_1(1/{\sqrt{n}}+{p}/{n^{3/2}})+\|\boldsymbol{\beta}^*\|_1^2\left({p}/{n^2}+{1}/{p}\right)=o(\|\boldsymbol{\beta}^*\|_2^2)$.
		%$=o(\|\boldsymbol{\beta}^*\|_2^2)$.
		%\item[($\text{ii}$)] 
		%$s^2{p(\log p)}=o({n})$.
		%$s/{\sqrt{n}}+s{p}/{n^{3/2}}=o(1)$.
		%\end{itemize}
	\end{assumption}
	
	%Here $s=|\mathcal{S}^*|$ is the cardinality of $\mathcal{S}^*=\left\{j \in [p]: \beta_{j}^* \neq 0\right\}$.  
	%Assumption \ref{Power1 assumption} imposes a sparsity condition under the alternative hypothesis.
	
	Assumption \ref{Power1 assumption} makes some restriction on
	$\|\boldsymbol{\beta}^*\|_1, \|\boldsymbol{\beta}^*\|_2, n$ and $p$. Denote $\delta={\boldsymbol{\beta}}^{*\top}\boldsymbol{\Sigma}_{\boldsymbol{u}}^2{\boldsymbol{\beta}}^{*}$.
	%\mathbb{E}\{({\boldsymbol{u}}_i^{\top}{\boldsymbol{\beta}}^*)({\boldsymbol{u}}_j^{\top}{\boldsymbol{\beta}}^*){\boldsymbol{u}}_i^{\top}{\boldsymbol{u}}_j\}=\sum_{j=1}^p\sum_{k',k''=1}^p{\boldsymbol{\beta}}_{k'}^*{\boldsymbol{\beta}}_{k''}^*\mathbb{E}(U_{ik'}U_{ij})\mathbb{E}(U_{ik''}U_{ij})$.
	We define the following parameter space for $H_1$,
	\begin{align} \label{parameter space1 for H1}
		\boldsymbol{\Omega}^{(1)}=\left\{\boldsymbol{\beta}^* \in \mathbb{R}^p: C_{n,p} =o(\delta)\right\},
	\end{align}
	where $C_{n,p}= {1}/{\sqrt{p}}+{1}/{\sqrt{n}}+{\sqrt{p}}/{n}+{p}/{n^{3/2}}$.
	This space corresponds to the dense alternative hypotheses. Here we provide a toy example. Suppose that each element of $\boldsymbol{\beta}^*$ is set as $0<c<\infty$. Then the Assumption \ref{Power1 assumption} and the existence of the space $\boldsymbol{\Omega}^{(1)}$ holds when $p=o(n^{3/2})$. While if we set $c=1/\sqrt{p}$, we require $p=o(n)$. These two settings correspond to strong signal setting and weak signal setting, respectively. 
	%where  $C_{n,p}=p(\log p)^5(\log n)^{7/2}/n^2+\sqrt{p}(\log p)^4(\log n)^{9/2}/n^{3/2}+(\log p)^2(\log n)^{4}/n+(\log p)(\log n)^{9/2}/\sqrt{np}$, and $c_0$ is a  positive constant.

	\begin{theorem} (Power analysis) \label{power1 analysis} Suppose that the conditions in Theorem \ref{sum statistic normal distribution} and Assumption \ref{Power1 assumption}  hold. 
		%In addition, we assume that  $\|{\boldsymbol{u}}^{\top}{\boldsymbol{\beta}}^*\|_{\psi_2} \leq C$ for  some constant $ C$. 
		For the parameter ${\boldsymbol{\beta}}^* \in \boldsymbol{\Omega}^{(1)}$, we have 
		\begin{align}
			\frac{{n}T_n}{\sigma_{\varepsilon}^2\sqrt{2\mbox{tr}(\boldsymbol{\Sigma}^2_{\boldsymbol{u}})}} \rightarrow \infty,  \ \text{as}\ \ (n,p) \rightarrow \infty.\notag 
		\end{align}
		%where $z(\alpha)$ is the upper-$\alpha$ quantile of $\mathrm{N}(0,1)$.
	\end{theorem}
	This theorem implies that for the dense alternative hypotheses in $\boldsymbol{\Omega}^{(1)}$, we can detect them with probability 1.

	In the following, we investigate the asymptotic distribution of the maximum-type test statistic $\widetilde S_n^2$. 
    Define $\boldsymbol{V}=(V_{ij})_{1 \leq i \leq n, 1\leq j \leq p} \in \mathbb{R}^{n \times p}$,  with ${V}_{ij}=[\mbox{Var}\{(Y_i-{\boldsymbol{f}}_{i}^{\top}{\boldsymbol{\gamma}}^*) {U}_{ij}\}]^{-1/2}$ $\{(Y_i-{\boldsymbol{f}}_{i}^{\top}{\boldsymbol{\gamma}}^*) {U}_{ij}-\mathbb{E}(Y_i-{\boldsymbol{f}}_{i}^{\top}{\boldsymbol{\gamma}}^*) {U}_{ij}\}$.
It is evident that ${V}_{ij}$ 
  represents the standardized version of the variable $(Y_i-{\boldsymbol{f}}_{i}^{\top}{\boldsymbol{\gamma}}^*) {U}_{ij}$.
Now, we impose the following assumptions on the covariance matrix  $\boldsymbol{\Sigma}^*=\mathbb{E}(\boldsymbol{v}\boldsymbol{v}^{\top})=({\Sigma}_{jk}^*)_{1 \leq j,k \leq p}$,  where $\boldsymbol{v}$ denotes a row of  $\boldsymbol{V}$.
   \iffalse  Let ${S}_{n,j}^*=n^{-1/2}\sigma_j^{\natural-1}\sum_{i=1}^n(Y_i-{\boldsymbol{f}}_{i}^{\top}{\boldsymbol{\gamma}}^*) {U}_{ij}$, where $\sigma_j^{\natural2}=\mathbb{E}\{U_{ij}^2(Y_i-{\boldsymbol{f}}_{i}^{\top}{\boldsymbol{\gamma}}^*)^2\}$.
    It could be shown that 
    
    Let  $\boldsymbol{\Lambda}^{\natural}=(\rho_{jk})_{1\leq j,k \leq p}$ be the covariance matrix of $(Y_i-{\boldsymbol{f}}_{i}^{\top}{\boldsymbol{\gamma}}^*) \boldsymbol{u}_{i}$, and  $\boldsymbol{\mathcal{R}}=\{{\mbox{diag}}^{-1/2}\left(\boldsymbol{\Sigma}^{\natural}\right)\}\boldsymbol{\Sigma}^{\natural}\{{\mbox{diag}}^{-1/2}\left(\boldsymbol{\Sigma}^{\natural}\right)\}=(\rho_{jk})_{1 \leq j,k \leq p}$.\fi
    For $0<r<1$, let $\mathcal{V}_{j}(r)=\left\{1\leq k\leq p:\left|{\Sigma}_{jk}^*\right|>r\right\}$ be the set of indices $k$ such that $V_{ik}$ is highly correlated (whose correlation $> r$) with $V_{ij}$ for a  given $j \in [p]$ and $i \in [n]$. Define $\mathcal{W}(r)=\{1 \leq j \leq p: \mathcal{V}_{j}(r) \neq \varnothing\}$ such that $\forall j \in \mathcal{W}(r)$, $V_{ij}$ is highly correlated with some other variables of $\boldsymbol{v}_i$.  For any $\tau >0$, let $s_j(\tau)=|\mathcal{V}_{j}\{(\log p)^{-1-\tau}\}|, \ j \in [p]$ be the number of indices $k$ in the set $\mathcal{V}_{j}\{(\log p)^{-1-\tau}\}$.
	
	\begin{assumption}\label{max statistic assumption}
		We make the following assumptions. 
		\begin{itemize} 
			%\item[($\text{i}$)]   There exists a positive constant $C$ such that $1/C \leq \Sigma^{*}_{jj} \leq C$ for $j=1,\ldots,p$.
            %Assume that $\inf_{1\leq j\leq p}\Sigma_{\boldsymbol{u},jj}\geq c$ for some positive constant $c$.
			%\item[($\text{ii}$)] 
			%There exists some constant $\varrho\in(0,1)$, such that $\left|\rho_{\boldsymbol{u},jk}\right|\leq \varrho$ for all $j\neq k$.
             \item[($\text{i}$)] There exists a constant $0 < r_0 <1$, $\left|\mathcal{W}(r_0)\right|=o(p)$.
             \item[($\text{ii}$)] There exists a constant $\tau_0 >0$, $\max_{j \in [p]}s_j(\tau_0)=o(p^{\kappa})$, for all $\kappa >0$.
						%Suppose that $\delta_p$ and $\varsigma_p$ are positive constants with $\delta_p=o\left\{(\log p)^{-1}\right\}$ and $\varsigma_p=o(1)$ as $p\to\infty$. For any $1\leq j\leq p$, define $\mathcal{V}_{j}(\delta_p)=\left\{1\leq k\leq p:\left|\rho_{\boldsymbol{u},jk}\right|>\delta_p\right\}$ and $\boldsymbol{C}_{p}(\delta_p)=\left\{1\leq j\leq p:\left|\mathcal{V}_{j}(\delta_p)\right|>p^{\varsigma_p}\right\}$. 
			%Assume that  $\max_{j \in [p]}\left|\mathcal{V}_{j}(\delta_p)\right|=o(p^{\kappa})$, for $\kappa >0$, and $\left|\boldsymbol{C}_{p}(\delta_p)\right|=o(p)$.
		\end{itemize}
	\end{assumption}
%Assumption \ref{max statistic assumption} (i) has been utilized in \cite{cao2018two, li2024power} to ensure that variances remain bounded away from zero and infinity. 
Assumption \ref{max statistic assumption} (i)  constrains the dependence structure among the components of $\boldsymbol{v}$. Similar condition was also considered in \cite{feng2022asymptotic, li2024power}.  Following \cite{tony2014two} and \cite{cao2018two}, to derive the limiting distribution of the maximum test statistic, we can instead impose assumptions  $\max_{1 \leq  j \neq k \leq p}|\Sigma^*_{jk}| \leq r_1$ and  $\max_{j \in [p]}\sum_{k=1}^p\Sigma^{*2}_{jk} \leq r_2$, for
constants $0<r_1<1$ and $r_2>0$. Notably, Assumption \ref{max statistic assumption} (ii) encompasses these as special cases when  $s_j(1)=C(\log p)^2$ \citep{li2024power}.

	\iffalse Assumption \ref{max statistic assumption} (i) has been  employed   in \cite{cao2018two, li2024power} to bound the variances away
	from zero and infinity. Assumption \ref{max statistic assumption} (ii)   imposes restrictions on the dependence structure among the components of $\boldsymbol{u}$, a condition also considered in \cite{feng2022asymptotic, li2024power}. Following \cite{cao2018two} and \cite{tony2014two}, to derive the limiting distribution of the maximum test statistic, we can also impose assumptions $\max_{1 \leq  j \neq k \leq p}|\Sigma^*_{jk}| \leq r_1 \leq 1$ and  $\max_{j \in [p]}\sum_{k=1}^p\Sigma^{*2}_{jk} \leq r_2$, for
constants $0<r_1<1$ and $r_2>0$. Notably, Assumption \ref{max statistic assumption} (iii) encompasses these as special cases when  $s_j(1)=C(\log p)^2$. \fi
	
	\begin{theorem}\label{max statistic theorem}
		Suppose that Assumptions \ref{factorassumption}-\ref{error sub-Gaussian assumption} and Assumption \ref{max statistic assumption} hold. In addition, we assume that $\{\log (np)\}^{7}=o(n)$ and $\sqrt{n}(\log p)(\log n)=o(p)$.    Under the null hypothesis $H_0$, for any $z\in\mathbb R$,  we have 
		%\begin{align}
		%\sup_{z\in\mathbb R} \left|\Pr\left(\tilde{S}_n -2\log p+\log\log p \leq z\right)- \exp\left\{-\frac{1}{\sqrt\pi}\exp\left(-\frac{z}{2}\right)\right\}\right|=o(1),
		%\end{align}
		\begin{align}
			\sup_{{z}\in\mathbb R} \left|\Pr\left[\tilde{S}_n^2 -2(\log p) +\{\log( \log p)\}\leq z\right] -\exp\left\{-\frac{1}{\sqrt\pi}\exp\left(-\frac{z}{2}\right)\right\}\right|=o(1). \notag  
		\end{align}
	\end{theorem}

	Compared with \cite{beyhum2023tuning}, we now derive the asymptotic distribution of the maximum-type test
	statistic $\widetilde S_n^2$. In addition,  we  allow the dimension $p$ to be in exponential order of the sample size. While in \cite{beyhum2023tuning}, the dimension is in polynomial order of the sample size. This is reasonable since here we make sub-Gausian assumption. To determine the critical value of $\tilde{S}_n^2$, the above theorem enables us to avoid estimating the high-dimensional parametric vector $\boldsymbol{\beta}^*$. While this is required for the procedure in \cite{beyhum2023tuning}. 
	
	Now we come to investigate the power performance of the maximum-type test
	statistic $\widetilde S_n^2$. Denote %$\sigma_j^{\natural2}=\mathbb{E}\{U_{ij}^2(Y_i-{\boldsymbol{f}}_{i}^{\top}{\boldsymbol{\gamma}}^*)^2\}$ and   
    $\delta_j^*=\sqrt{n}\sum_{k=1}^{p}\beta_k^*\mathbb{E}(U_{ij}U_{ik})/\sigma_j^{\natural}$, with $\sigma_j^{\natural2}=\mathbb{E}\{U_{ij}^2(Y_i-{\boldsymbol{f}}_{i}^{\top}{\boldsymbol{\gamma}}^*)^2\}$. We define the following parameter space for $H_1$,
	\begin{align} \label{parameter space for H1}
		\boldsymbol{\Omega}^{(2)}(c_0)=\left\{\boldsymbol{\beta}^* \in \mathbb{R}^p: \max\limits_{j \in [p]}|\delta_j^*|\geq 
		\sqrt{c_0 {\log p}}\right\},
	\end{align}
	where $c_0$ is a positive constant.
	Furthermore,  if we assume that $\boldsymbol{\Sigma}_{\boldsymbol{u}}$ is a diagonal matrix, implying that the components of $\boldsymbol{u}$ are uncorrelated—an assumption commonly made in factor analysis \citep{BaiandNg2002, Bai2003}—then $\delta_j^*$ can be simplified to $\delta_j^*=\sqrt{n}\beta_j^*\mathbb{E}(U_{ij}^2)/\sigma_j^{\natural}$.
	This space corresponds to the sparse alternative hypotheses. 
\iffalse		
Define $\boldsymbol{V}=(V_{ij})_{1 \leq i \leq n, 1\leq j \leq p} \in \mathbb{R}^{n \times p}$,  with ${V}_{ij}=[{\sigma_{j}^{ \natural2}-\{\sum_{k=1}^{p}\beta_k^*\mathbb{E}(U_{ij}U_{ik})\}^2}]^{-1/2}\{(Y_i-{\boldsymbol{f}}_{i}^{\top}{\boldsymbol{\gamma}}^*) {U}_{ij}-\sqrt{n}\sum_{k=1}^{p}\beta_k^*\mathbb{E}(U_{ij}U_{ik})\}$.
It is evident that ${V}_{ij}$ 
  represents the standardized version of the variable $(Y_i-{\boldsymbol{f}}_{i}^{\top}{\boldsymbol{\gamma}}^*) {U}_{ij}$.
Now, we impose the following assumption on the covariance matrix  $\boldsymbol{\Sigma}^*=\mathbb{E}(\boldsymbol{v}\boldsymbol{v}^{\top})$,  where $\boldsymbol{v}$ denotes a row of  $\boldsymbol{V}$.

%let $\boldsymbol{z}=(z_1, \ldots, z_p)^{\top} \in \mathbb{R}^p$ be a zero-mean multivariate normal
% random vector such that $\boldsymbol{z}$   has the same covariance matrix as $\boldsymbol{\Sigma}^*=\mathbb{E}(\boldsymbol{v}\boldsymbol{v}^{\top})$, and $\boldsymbol{v}$ is the corresponding row of $\boldsymbol{V}$.  We make the following assumption. 
 \begin{assumption}  \label{covariance assumption}
 Suppose that 
     $\max_{1 \leq j \neq k \leq p}|\Sigma_{jk}^*| \leq r <1$ and $\max_{k \in [p]}\sum_{j=1}^p\Sigma_{jk}^{*2} \leq C_0$, where $C_0$ is a positive constant.
 \end{assumption}
 Assumption \ref{covariance assumption} imposes a mild constraint on the internal correlation of vector $(Y_i-{\boldsymbol{f}}_{i}^{\top}{\boldsymbol{\gamma}}^*) \boldsymbol{u}_{i}$, which has also been adopted in \cite{tony2014two}.
 \fi 
 \begin{assumption} \label{power2 assumption}
		The following holds:
		%\begin{itemize} 
		%\item[($\text{i}$)] 
		$\|\boldsymbol{\beta}^*\|_1\left\{ (\log p)^2/\sqrt{n}+\sqrt{n}(\log p)/p\right\}=o(1)$.
		%\item[($\text{ii}$)] $s\left\{\frac{(\log p)^{7/4}}{\sqrt{n}}\right\}=o(1)$.
		%\end{itemize}
	\end{assumption}
	The above assumption is similar to the Assumption 5 (ii) in \cite{beyhum2023tuning}.

	\begin{theorem} (Power analysis.) \label{Power2 analysis}
		Suppose that the conditions in Theorem \ref{max statistic theorem} and Assumption  \ref{power2 assumption} hold. 
        %In addition, we assume that there exist some positive  constants $c<C$ such that $\sigma_j^{\natural2} \geq c$, and  $C$ satisfies that $\|{\boldsymbol{u}}^{\top}{\boldsymbol{\beta}}^*\|_{\psi_2} \leq C$.
		%and  $\|U_{ij}(Y_i-{\boldsymbol{f}}_{i}^{\top}{\boldsymbol{\gamma}}^*)\|_{\psi_1} \leq C$ for all $1 \leq j \leq p$.
		%If $(\log n)^4= o( p)$ and $(\log p)^4(\log n)^2=o(n)$,
		%$\sqrt{n}(\log p)^{3/2}(\log n) \ll p$ and  $\{\log (pn)\}^7/n=o(1)$, 
		For the test statistic $\tilde{S}_n^2$, we have for some $c_1 >0$, 
		\begin{align}
			\lim\limits_{(n,p)\rightarrow \infty} \inf\limits_{\boldsymbol{\beta}^* \in  
				\boldsymbol{\Omega}^{(2)}(1+c_1)} \Pr\left\{ \tilde{S}_n^2 
			> c(\alpha)\right\}=1, \notag 
		\end{align}
		where $c(\alpha)$ is defined in Section \ref{Methodology}.
	\end{theorem}

	\iffalse
	\begin{assumption}
		Denote ${\boldsymbol{\xi}}_i={\boldsymbol{u}}_i{\varepsilon}_i, \ i=1,\ldots,n$, which satisfies the following conditions:
		%and $\boldsymbol{\Sigma}_{\boldsymbol{u}}^\natural=\sigma^2_{\varepsilon}\mathbf\Sigma_{\boldsymbol{u}}$. 
		\begin{itemize} 
			\item[($\textit{i}$)] there exists a constant $C$ such that, for any $\boldsymbol{\alpha} \in \mathbb{R}^p$, 
			\begin{align}
				(\mathbb{E}|\boldsymbol{\alpha}^{\top}{\boldsymbol{\xi}}|^4)^{1/4} \leq C(\mathbb{E}|\boldsymbol{\alpha}^{\top}{\boldsymbol{\xi}}|^2)^{1/2}. \notag
			\end{align}
			\item[($\textit{ii}$)] $\mathbb{E}(\max\limits_{1 \leq k \leq p}|\xi_{ik}|^3)\leq M^3(\log p)^{3/2}$, with $(\log p)^{10}M^6=o(n)$.
		\end{itemize}
	\end{assumption}\fi
	
	In the following, we aim to establish the asymptotic independence of the quadratic-type test statistic $M_n$ and the maximum-type test statistic $\widetilde S_n^2$ under the null hypothesis.  
    \iffalse We  make the following assumption.
    \begin{assumption} \label{independence assumption}
    There exists a constant $\tau_0 >0$, $\max_{j \in [p]}s_j(\tau_0)=o(p^{\kappa})$, for all $\kappa >0$.
    \end{assumption}
% Assumption \ref{independence assumption} is also considered in \cite{li2024power}, and it  is milder compared to the condition in \cite{cao2018two}.
 Assumption \ref{independence assumption}, also considered in \cite{li2024power}, imposes a weaker condition than that in \cite{cao2018two}, making it more broadly applicable.
\fi
    
	%Denote $\boldsymbol{\Sigma}^\natural=({\Sigma}_{ij}^\natural)_{p \times p}$.  
	%Moreover, for $1 \leq j \leq p$ and $0< \alpha < 1$,  define $s_{j}(\alpha)=\text{card}[\mathcal{V}_{j}\{(\log p)^{-1-\alpha}\}]$, where  $\mathcal{V}_{j}(\delta_p)=\left\{1\leq k\leq p:\left|\rho_{\boldsymbol{u},jk}\right|>\delta_p\right\}$.
	%Define 
	%\begin{align}
	%\mathcal{\boldsymbol{V}}_i(\Sigma^{\natural},r)=\left\{1 \leq j \leq p: \left|\frac{\Sigma_{ij}^{\natural}}{\sqrt{\Sigma_{ii}^{\natural}}\Sigma_{jj}^{\natural}}\right|\right\}
	%\end{align}

	\iffalse
	\begin{assumption} \label{u independent assumption}
		Assume that all elements of the vector $\boldsymbol{u}$ are mutually independent.
	\end{assumption} \fi
	\begin{theorem} \label{sum and max independent theorem}
		Suppose that the conditions in Theorems \ref{sum statistic normal distribution} and \ref{max statistic theorem} hold. Under the null hypothesis $H_0$, the following property holds:
		\begin{align}
			\Pr\left[M_n \leq x, \tilde{S}_n^2-2(\log p)+\{\log(\log p)\} \leq y\right]\xrightarrow{{d}} \Phi(x)\cdot F(y) \notag 
		\end{align}
		for any $x,y \in \mathbb{R}$, as $n,p \rightarrow  \infty$. Here, $\Phi(\cdot)$ and $F(\cdot)$ are the  cumulative distribution functions  of standard normal distribution and Gumbel distribution, respectively. 
		%$M_n$ and $\tilde{S}_n$ are asymptotically independent, where 
		%$\widehat\sigma_{j}^2$ is a consistent estimator of $\sigma^2_{\varepsilon}\Sigma_{\boldsymbol{u},jj}$
	\end{theorem}
	
	The above important result implies that we can combine the strength of both $M_n$ and $\widetilde S_n^2$. Actually based on the above Theorem, we can introduce an adaptive test procedure $F_n$, which is defined in section \ref{Methodology}. From the above Theorem, we can obtain the following result, which implies that the adaptive test procedure $F_n$ can control the size asymptotically. 
	
	\begin{corollary}\label{Asymptotic Size theorem}
		Under the conditions in Theorem \ref{sum and max independent theorem} and the null hypothesis, we have
		\begin{align}
			\Pr(F_n \geq \chi_{\alpha})\rightarrow \alpha, \ \text{as} \ n,p \rightarrow \infty. \notag  
		\end{align}
        Here, $\chi_{\alpha}$ is the upper $\alpha$-quantile of the
	$\chi_4^2$ distribution. 
	\end{corollary}
	
	From Theorem \ref{power1 analysis} and Theorem \ref{Power2 analysis}, we can further obtain the following result. 
	\begin{corollary} \label{Asymptotic power theorem}
		Suppose that the conditions in Theorem \ref{power1 analysis} and \ref{Power2 analysis} hold. We then have
		\begin{align}
			\lim_{(n,p)\rightarrow \infty}\inf_{\boldsymbol{\beta}^* \in \boldsymbol{\Omega}^{(1)}\cup  \boldsymbol{\Omega}^{(2)}(c)} \Pr(F_n \geq \chi_{\alpha})=1. \notag
		\end{align}
	\end{corollary}
	The above Corollary implies that our introduced adaptive test statistic $F_n$  is powerful to detect signals under either sparse or dense alternative hypotheses.

	\section{Simulation studies}
	\label{simulations}
	% We generate the samples under three $(n,p)$-configurations: $(n,p) \in \{(200,100),(200,200),$ $(200,500)\}$, and choose  $K=2$.  
	% In addition, we generate every row of $\boldsymbol{F} \in \mathbb{R}^{n \times K}$ from $\mathrm{N}(\boldsymbol{0}, \boldsymbol{I}_K)$, and every row of $\boldsymbol{U} \in \mathbb{R}^{n \times p}$ follows from $\mathrm{N}(\boldsymbol{0}, \boldsymbol{\Sigma}_{\boldsymbol{u}})$. The $\boldsymbol{\Sigma}_{\boldsymbol{u}}$ are set to two different structures. The first is an identity matrix $\boldsymbol{I}_p$, and the second is an AR(1) matrix, $\boldsymbol{\Sigma}_{\boldsymbol{u}}^{\text{AR}}=(\Sigma_{u,jk})_{1 \leq j,k \leq p}$ with $\Sigma_{u,jk}=0.3^{|j-k|}$. Here  the entries of $\boldsymbol{B}$ are  generated from the uniform  distribution $\mathrm{Unif}(-1,1)$. We set ${\bX}=\boldsymbol{F}\boldsymbol{B}^{\top}+\boldsymbol{U}$.
	
	We generate the samples under three \((n, p)\)-configurations: \((n, p) \in \{(200,100), (200,200), $ $(200,500)\}\), and set \(K = 2\).  
	Additionally, each row of \(\boldsymbol{F} \in \mathbb{R}^{n \times K}\) is drawn from \(\mathrm{N}(\boldsymbol{0}, \boldsymbol{I}_K)\), and each row of \(\boldsymbol{U} \in \mathbb{R}^{n \times p}\) is sampled from \(\mathrm{N}(\boldsymbol{0}, \boldsymbol{\Sigma}_{\boldsymbol{u}})\). The covariance matrix \(\boldsymbol{\Sigma}_{\boldsymbol{u}}\) is considered in two different structures: the first is an identity matrix \(\boldsymbol{I}_p\), and the second is an AR(1) matrix, \(\boldsymbol{\Sigma}_{\boldsymbol{u}}^{\text{AR}} = (\Sigma_{u,jk})_{1 \leq j,k \leq p}\), where \(\Sigma_{u,jk} = 0.3^{|j-k|}\).  
	The entries of \(\boldsymbol{B}\) are independently drawn from the uniform distribution \(\mathrm{Unif}(-1,1)\). Finally, we define \(\bX = \boldsymbol{F} \boldsymbol{B}^{\top} + \boldsymbol{U}\).  
	
Firstly, we conduct simulation studies across different sparsity levels under three \((n, p)\)-configurations.  We compare four test procedures: the maximum-type tests proposed in this paper ($\tilde{S}_n$) and in \cite{beyhum2023tuning} ($S_n^B$), as well as the quadratic-type test  ($M_n$) and the Fisher adaptive test  ($F_n$). The ${S}_n^B$ test in \cite{beyhum2023tuning}  considered the test statistic $S_{n}=\max_{1\leq j\leq p}|n^{-1/2}\sum_{i=1}^n\widehat U_{ij} (Y_i-\hat{\boldsymbol{f}}_i^{\top}\hat{\boldsymbol{\gamma}})|$ but employed Gaussian multiplier bootstrap for critical value determination.   We consider three alternative settings to assess the comparative performance of these methods: (i) the fully sparse case ($s=1$),  (ii) the moderately dense case ($s=10$), and (iii) the fully dense case ($s=p$). Here $s$ is the number of nonzero elements of $\boldsymbol{\beta}^*$. We generate  \(\beta_j^* = \omega\) for \(1 \leq j \leq s\) and \(\beta_j^* = 0\) otherwise, with $\omega$ ranging from 0 to 0.3. When $\omega=0$, the null hypothesis holds, and the results reflect the empirical size; otherwise, they correspond to the empirical power.  
    Additionally, we set \(\boldsymbol{\gamma}^* = (0.5,0.5)^{\top}\). The response vector is given by  
	\[
	\boldsymbol{Y} = \boldsymbol{F} \boldsymbol{\gamma}^* + \boldsymbol{U} \boldsymbol{\beta}^* + \boldsymbol{\varepsilon},
	\]  
	where each entry of \(\boldsymbol{\varepsilon} \in \mathbb{R}^n\) is independently drawn from \(\mathrm{N}(0, 0.25)\).  For the test $S_n^B$, we implement it by the R package `FAS', set the grid  size to 100 and choose an equidistant grid of values for tuning parameter,  using 1000 bootstrap replications to obtain the critical value. Each test is evaluated based on 2000 repetitions at the significance level  \(\alpha = 0.05\). 
    
The simulation results, presented in Figures~\ref{p_100_sfix}–\ref{p_500_sfix}, lead to the following conclusions. We first note that all tests can control their type-I error rates well. Under all three  \((n, p)\)-configurations and two covariance settings, when $\boldsymbol{\beta}^*$ is fully sparse, the power curves of the maximum-type tests $\tilde S_n, S_n^B$ and Fisher adaptive test $F_n$ rapidly converge to 1 at a comparable rate, significantly faster than that of the quadratic-type test $M_n$. In the moderately dense case, all  tests exhibit similar power growth rates, with the Fisher adaptive test and quadratic-type test slightly outperforming the maximum-type tests. Conversely, in the fully dense scenario, the quadratic-type test and Fisher adaptive test quickly reach a power of 1, both significantly faster than the maximum-type tests. Moreover, the proposed maximum-type test $\tilde{S}_n$ demonstrates superior power compared to ${S}_n^B$ in most cases. These results indicate that the maximum-type tests perform best in sparse settings but loses effectiveness in dense cases, whereas the quadratic-type test excels in dense settings but is less powerful in sparse cases.  The proposed Fisher adaptive test, on the other hand, combines the strengths of both tests. It performs as well as the maximum-type tests when the number of nonzero coefficients is small (sparse case) and achieves nearly the same power as the quadratic-type test when the number of nonzero coefficients increases (non-sparse case). Notably, all tests, except for the Fisher adaptive test, favor either the sparse or the non-sparse setting. Since it is often difficult to determine the true sparsity  in practice, the Fisher adaptive test, with its robustness across different sparsity levels, emerges as a more favorable choice compared to the competing  approaches.
\begin{center}
	 		Figures~\ref{p_100_sfix}--\ref{p_500_sfix} should be here.
	 	\end{center}

	In the following analysis, we further design the parametric vector $\boldsymbol{\beta}^*$  to ensure signal comparability across all sample size and dimensionality configurations.  
	To achieve this, we set the nonzero entries of \(\boldsymbol{\beta}^* \in \mathbb{R}^p\) to be equal such that $\|\boldsymbol{\beta}^*\|_2= 0.2$.
The empirical type-I error rate is assessed by setting sparsity \(s = 0\).  To evaluate empirical power under sparse alternatives, we consider the sparsity levels \(s = 1, 3, 5, 7, 9, 11\), while for dense alternatives, we set \(s = 50\%p, 60\%p, 70\%p, 80\%p, 90\%p, 100\%p\).   We generate \(\beta_j^* = \omega\), with $\omega=0.2/\sqrt{s}$ for \(1 \leq j \leq s\) and \(\beta_j^* = 0\) otherwise.  
	%To evaluate test performance under varying signal sparsity levels, we consider different proportions of nonzero entries, chosen from \(\{0.01p, 0.05p, 0.2p, 0.5p, 0.8p, p\}\), with their locations randomly selected. 
    As \(\boldsymbol{\beta}^*\) becomes denser, the individual signal strength of each element decreases. 
    We evaluate each test based on 2000 simulations at a significance level of \(\alpha = 0.05\) and report the empirical rejection percentages in Figures~\ref{p_100_fig}--\ref{p_500_fig}. 

	From Figures~\ref{p_100_fig}--\ref{p_500_fig}, we draw the following conclusions.  
	Across all covariance structures and data-generating scenarios, the  tests \(\tilde{S}_n\),  \(M_n\), \(F_n\) and $S_n^B$ effectively control the type-I error rates. The maximum-type tests \(\tilde{S}_n\) and $S_n^B$ exhibit comparable powers and follow similar trends across different sparsity settings. Additionally, as noted earlier, when the signal is highly sparse, the maximum-type tests exhibit greater power, whereas under denser alternatives, the quadratic-type test demonstrates superior power.  
	Specifically, under sparse alternatives,  when $s=1$ and $3$, the maximum-type tests outperform the quadratic-type test. However, when sparsity  $s$ ranges from 5 to 11, the quadratic-type test generally achieves higher power.  Under dense alternatives, the quadratic-type test consistently demonstrates higher power than the maximum-type tests. Furthermore, when the covariance matrix is the identity matrix \(\boldsymbol{I}_p\), the maximum-type tests experience a significant loss of power as the signal sparsity increases, as expected, while the power of the quadratic-type test remains relatively high. In scenarios with an AR(1) covariance structure, the power of the maximum-type tests declines at a slower rate but still remains lower than that of the quadratic-type test as sparsity increases.
    Finally, as the dimensionality $p$ increases from 100 to 500, the power of all tests declines to varying degrees, which is consistent with expectations.

	% Adaptive tests offer a safeguard against incorrectly assumed sparsity. Across all covariance structures and data-generating scenarios, the Fisher adaptive test demonstrates robustness to the true underlying sparsity of the signal. Specifically, in the sparsest settings, the maximum-type test is the most effective. However, we observe that the Fisher adaptive test not only exhibits greater power than the quadratic-type test but also matches the power of the maximum-type test. As the signal density increases, the maximum-type test experiences a sharp decline in power. In contrast, the adaptive tests remain unaffected, benefiting from the strong performance of the quadratic-type test. In the densest settings, the Fisher adaptive test exhibits power comparable to that of the quadratic-type test and, in certain sparse covariance settings, even outperforms it. 
	
	Adaptive test provides a safeguard against different sparsity levels. Across all covariance structures and data-generating scenarios, the Fisher adaptive test exhibits robustness to the underlying sparsity of the signal. Specifically, in the sparsest settings, the maximum-type tests are the most effective. However, the Fisher adaptive test not only achieves greater power than the quadratic-type test but also matches the power of the maximum-type test. As the signal density increases, the maximum-type tests undergo a sharp decline in power. In contrast, adaptive test remains unaffected, leveraging the strong performance of the quadratic-type test. In the densest settings, the Fisher adaptive test attains power comparable to that of the quadratic-type test and, in certain sparse covariance scenarios, even surpasses it.
    \begin{center}
	 		Figures~\ref{p_100_fig}--\ref{p_500_fig} should be here.
	 	\end{center}

As previously mentioned, \cite{beyhum2023tuning} determined the critical value of $S_n$  using Gaussian multiplier bootstrap method  and also estimated the high-dimensional parameter vector $\boldsymbol{\beta}^*$. In contrast, the asymptotic distribution of  $\tilde{S}_n^2$ is derived without requiring such an estimation, leading to a reduction in computational cost. Table~\ref{average computation time} compares the average computation time of tests $\tilde{S}_n$ and $S_n^B$ under the fully dense setting, ${\beta}_j^*=0.2/\sqrt{p}, \ j=1,\ldots,p$. The results indicate that, under identical settings, the test $\tilde{S}_n$ requires significantly less computation time than $S_n^B$. Furthermore, as the parameter dimension $p$ increases from 100 to 500, the computational cost for both tests rises accordingly, which is expected.
 \begin{center}
	 		Table~\ref{average computation time} should be here.
	 	\end{center}

	\section{Real data analysis}
	\label{Real data analysis section}
	In this Section, we employ the FRED-MD macroeconomic dataset \citep{mccracken2016fred} to evaluate and compare the performance of the maximum-type tests proposed in this paper ($\tilde{S}_n$) and in \cite{beyhum2023tuning} ($S_n^B$), as well as the quadratic-type test  ($M_n$) and the Fisher adaptive test  ($F_n$).  This dataset is widely used in macroeconomic research and has also been analyzed by \cite{fanJ2023} and \cite{beyhum2023tuning}.  It is publicly available at  \url{https://www.stlouisfed.org/research/economists/mccracken/fred-databases} \citep{mccracken2016fred}.
    It comprises 134 monthly U.S. macroeconomic variables, which capture various aspects of economic activity. As these variables are influenced by latent factors, they exhibit significant interdependencies \citep{fanJ2023}.
	
	Following the methodology proposed by \cite{mccracken2016fred},  we apply standard transformations to the original dataset. However, there exist significant structural breaks for many variables  around the 2008 financial crisis, leading to nonstationarity even after transformation. To address this issue, and in line with \cite{fanJ2023}, we conduct our analysis separately for two distinct time periods: February 1992 - October 2007 and August 2010 - February 2020, after carefully examining data availability and stationarity. 
	In our analysis, we select four variables: ``DPCERA3M086SBEA'', ``UEMP5TO14'', ``USTPU'', and ``AMDMUOx'' as response variables respectively, while treating the remaining variables as covariates. Here,   ``DPCERA3M086SBEA'' represents real personal consumption expenditures, a key measure of inflation-adjusted consumer spending on goods and services.   ``UEMP5TO14''  denotes the unemployment rate for individuals aged 5 to 14, reflecting labor market participation and employment conditions within this age group.   ``USTPU''  captures the unemployment rate of individuals uncertain about the duration of their unemployment, offering insights into labor market uncertainty and employment stability.   ``AMDMUOx''   represents manufacturers' unfilled orders (total), measuring the dollar value of orders received by manufacturers that remain uncompleted or unshipped, serving as an indicator of demand pressure in the manufacturing sector.
	
	We employ principal component analysis to estimate the latent  factors $\boldsymbol{f}$ and $\boldsymbol{u}$, and determine the number of factors using the eigenvalue ratio estimator. 
    %We then apply our testing procedures to test  $H_0$.
    The $p$-values of $\tilde{S}_n$,  $M_n$, $F_n$ and $S_n^B$  are provided in Table \ref{p-values of real data}. At the significance level of 0.05,   the results indicate that during the period February 1992 - October 2007, all four testing procedures reject the null hypothesis for ``UEMP5TO14'' and ``AMDMUOx'', indicating that the latent factor regression model is inadequate for these variables. However, for ``DPCERA3M086SBEA'' and ``USTPU'', the tests $\tilde{S}_n$, $M_n$, and $F_n$ reject the null hypothesis, whereas $S_n^B$ does not.
    In contrast, during the period August 2010 - February 2020, all four tests reject the null hypothesis for ``DPCERA3M086SBEA'' and ``USTPU'', again suggesting that the latent factor regression model fails to adequately capture their structure.  For ``UEMP5TO14'', none of the four tests reject the null hypothesis. As for ``AMDMUOx'', all tests except for $S_n^B$ reject the null hypothesis, consistent with the findings of \cite{beyhum2023tuning}.
    These results demonstrate that $\tilde{S}_n$, $M_n$, and $F_n$ exhibit greater power in detecting signals within the dataset compared to $S_n^B$. Moreover, they underscore the importance of incorporating the idiosyncratic component $\boldsymbol{u}$ into the regression model to more effectively capture the underlying data structure.
    \iffalse
     all three testing procedures reject the null hypothesis for ``AWOTMAN'', ``CLAIMSx'', and ``UMCSENTx'', suggesting that the latent factor regression model is inadequate for these variables. However, for ``BUSLOANS'', only the quadratic-type test rejects the null hypothesis but with relatively weak evidence. Although the Fisher adaptive test yields a smaller $p$-value than the maximum-type test, neither provides sufficient evidence to reject the null hypothesis.   
    For the period August 2010 - February 2020, all three tests reject the null hypothesis for ``AWOTMAN'' and ``BUSLOANS'', again indicating that the latent factor regression model fails to adequately describe these variables. In contrast, for ``CLAIMSx'' and ``UMCSENTx'', the quadratic-type test and Fisher adaptive test reject the null hypothesis, while the maximum-type test fails to do so. These findings highlight that the quadratic-type test and Fisher adaptive test are more powerful to detect signals within the dataset. Moreover,   incorporating the idiosyncratic component $\boldsymbol{u}$ into the regression model is essential for effectively capturing the underlying data structure. \fi
\begin{center}
	 		Table \ref{p-values of real data} should be here.
	 	\end{center}

	\iffalse \begin{table}[H]
		\small
		\renewcommand\arraystretch{1.1}
		\centering \tabcolsep 12pt \LTcapwidth 6in
		\caption{$p$-values of the maximum-type test, the quadratic-type test, and the Fisher adaptive test  for testing the adequacy of the latent factor regression models to explain ``AWOTMAN'', ``BUSLOANS'', ``CLAIMSx'', and ``UMCSENTx'' data in two different time periods.}
		\label{p-values of real data}
		\begin{threeparttable}
			\begin{tabular}{ccccc}
				\toprule
				Time Period & Data    & ${M}_n$   & $\tilde{S}_n$ & $F_n$    \\ \midrule
				\multirow{4}{*}{1992.02-2007.10} &AWOTMAN    & $8.04 \times 10^{-4}$ & $3.24 \times 10^{-3}$ & $3.61 \times 10^{-5}$ \\
				&BUSLOANS     & $4.57 \times 10^{-2}$ & $2.06 \times 10^{-1}$ & $5.34 \times 10^{-2}$ \\
				&CLAIMSx     & $1.06 \times 10^{-4}$ & $2.81 \times 10^{-2}$ & $4.09 \times 10^{-5}$ \\
				&UMCSENTx & $7.40 \times 10^{-6}$ & $1.32 \times 10^{-4}$ & $2.12 \times 10^{-8}$ \\  \hline
				\multirow{4}{*}{2010.08-2020.02} &AWOTMAN     & $4.06 \times 10^{-4}$ & $8.55 \times 10^{-3}$  & $4.71 \times 10^{-5}$\\
				&BUSLOANS     & $3.99 \times 10^{-7}$ & $1.57 \times 10^{-2}$  & $1.25 \times 10^{-7}$\\
				&CLAIMSx     & $8.73 \times 10^{-3}$ & $1.48 \times 10^{-1}$  & $9.87\times 10^{-3}$\\
				&UMCSENTx & $4.96 \times 10^{-4}$ &$9.27 \times 10^{-2}$ &$5.05 \times 10^{-4}$ \\
				\bottomrule
			\end{tabular}
		\end{threeparttable}
	\end{table} \fi

\section{Conclusions and discussions}
\label{Conclusions and discussions section}
In this paper, we investigate the adequacy testing problem of high-dimensional factor-augmented regression model. We introduce a novel quadratic-type test statistic which is powerful under dense alternative hypotheses. We further propose an adaptive test procedure which is powerful to detect both sparse and dense alternative hypotheses. We establish the asymptotic normality of the proposed quadratic-type test statistic and asymptotic independence of the newly introduced quadratic-type test statistic and an existing maximum-type test. In practice, the introduced Fisher adaptive test procedure is recommended since the true sparsity structure is often difficult to determine and the Fisher adaptive test procedure performs well across different sparsity levels.
	
To derive the asymptotic normality of the quadratic-type test statistic, we make some restrictions on the dimension $p$. This is necessary as we have to control the estimation errors from $\hat{\boldsymbol{u}}_i-{\boldsymbol{u}}_i$. It would be interesting to relax the restriction on the dimension $p$. Further outliers and heavy-tailed data often exist in practice. It is then important to develop robust test procedures to determine whether a factor regression model is adequate. We will explore these issues in near future.

	{ %\baselineskip 17pt
		\bibliographystyle{apalike}
		\bibliography{bibliography0606}
	}
	
\newpage
	\section*{Tables}
	\setcounter{table}{0} % 重置表格计数器
	\renewcommand{\thetable}{\arabic{table}} % 确保表格编号为阿拉伯数字格式

    \begin{table}[H]
		\small
		
		\renewcommand\arraystretch{0.9}
		\centering \tabcolsep 12pt \LTcapwidth 6in
		\caption{The average computation time (Unit: second). The $\tilde{S}_n$ and $S_n^B$ denote the maximum-type tests proposed in this paper and that in \cite{beyhum2023tuning}, respectively.} \label{average computation time}
		\begin{threeparttable}
			\begin{tabular}{>{\centering\arraybackslash}p{1.5cm}  >{\centering\arraybackslash}p{2cm}  >{\centering\arraybackslash}p{2cm}  >{\centering\arraybackslash}p{2cm}  >{\centering\arraybackslash}p{2cm}}
				\toprule
				\multirow{2}{*}{$(n,p)$}  & \multicolumn{2}{c}{$\boldsymbol{\Sigma}_{\boldsymbol{u}}=\boldsymbol{I}_p$} & \multicolumn{2}{c}{$\boldsymbol{\Sigma}_{\boldsymbol{u}}=\text{AR}(1)$} \\
				               &   $\tilde{S}_n$      & $S_n^B$   &  $\tilde{S}_n$     &$S_n^B$   \\ \hline
				\multirow{1}{*}{$(200,100)$} &            0.004     & 0.067 &  0.004       &0.083   \\ 
				\multirow{1}{*}{$(200,200)$} &            0.006     & 0.148  & 0.006       &0.157    \\ 
				\multirow{1}{*}{$(200,500)$} &            0.030     & 0.617  & 0.034       &0.813  
				 \\  \bottomrule
			\end{tabular}
		\end{threeparttable}
	\end{table}

     \begin{table}[H]
		\small
		\renewcommand\arraystretch{0.9}
		\centering \tabcolsep 12pt \LTcapwidth 6in
		\caption{The $p$-values of the maximum-type tests proposed in this paper ($\tilde{S}_n$) and in \cite{beyhum2023tuning} ($S_n^B$), as well as the quadratic-type test  ($M_n$) and the Fisher adaptive test  ($F_n$) for testing the adequacy of the latent factor regression models to explain ``DPCERA3M086SBEA'', ``UEMP5TO14'', ``USTPU'', and ``AMDMUOx'' data in two different time periods.}
		\label{p-values of real data}
		\begin{threeparttable}
			\begin{tabular}{cccccc}
				\toprule
				Time Period & Data    & $\tilde{S}_n$   & ${M}_n$ & $F_n$ & $S_n^B$  \\ \midrule
				\multirow{4}{*}{1992.02-2007.10} &DPCERA3M086SBEA    & $<10^{-3}$ & $<10^{-3}$ & $<10^{-3}$ & $0.1580$\\
				&UEMP5TO14     & $<10^{-3}$ & $0.0091$ & $<10^{-3}$ & 0.0415 \\
				&USTPU     & $<10^{-3}$ & $<10^{-3}$ & $<10^{-3}$ &0.9990 \\
				&AMDMUOx &$<10^{-3}$  & $<10^{-3}$ & $<10^{-3}$ & $0.0025$ \\  \hline
				\multirow{4}{*}{2010.08-2020.02} &DPCERA3M086SBEA     &$<10^{-3}$  & $<10^{-3}$  & $<10^{-3}$ &$0.0105$ \\
				&UEMP5TO14     & $0.0808$& $0.2443$   & $0.0972$ &$0.9890$ \\
				&USTPU     & $<10^{-3}$ &  $<10^{-3}$ & $<10^{-3}$ &0.0005 \\
				&AMDMUOx & $<10^{-3}$ &$<10^{-3}$ &$<10^{-3}$ &$0.9990$ \\
				\bottomrule
			\end{tabular}
		\end{threeparttable}
	\end{table}

    \newpage
	\section*{Figures}
		\setcounter{figure}{0}
	\renewcommand{\thefigure}{\arabic{figure}}

   \begin{figure}[H]%
		\centering
		\includegraphics[width=1.1\textwidth]{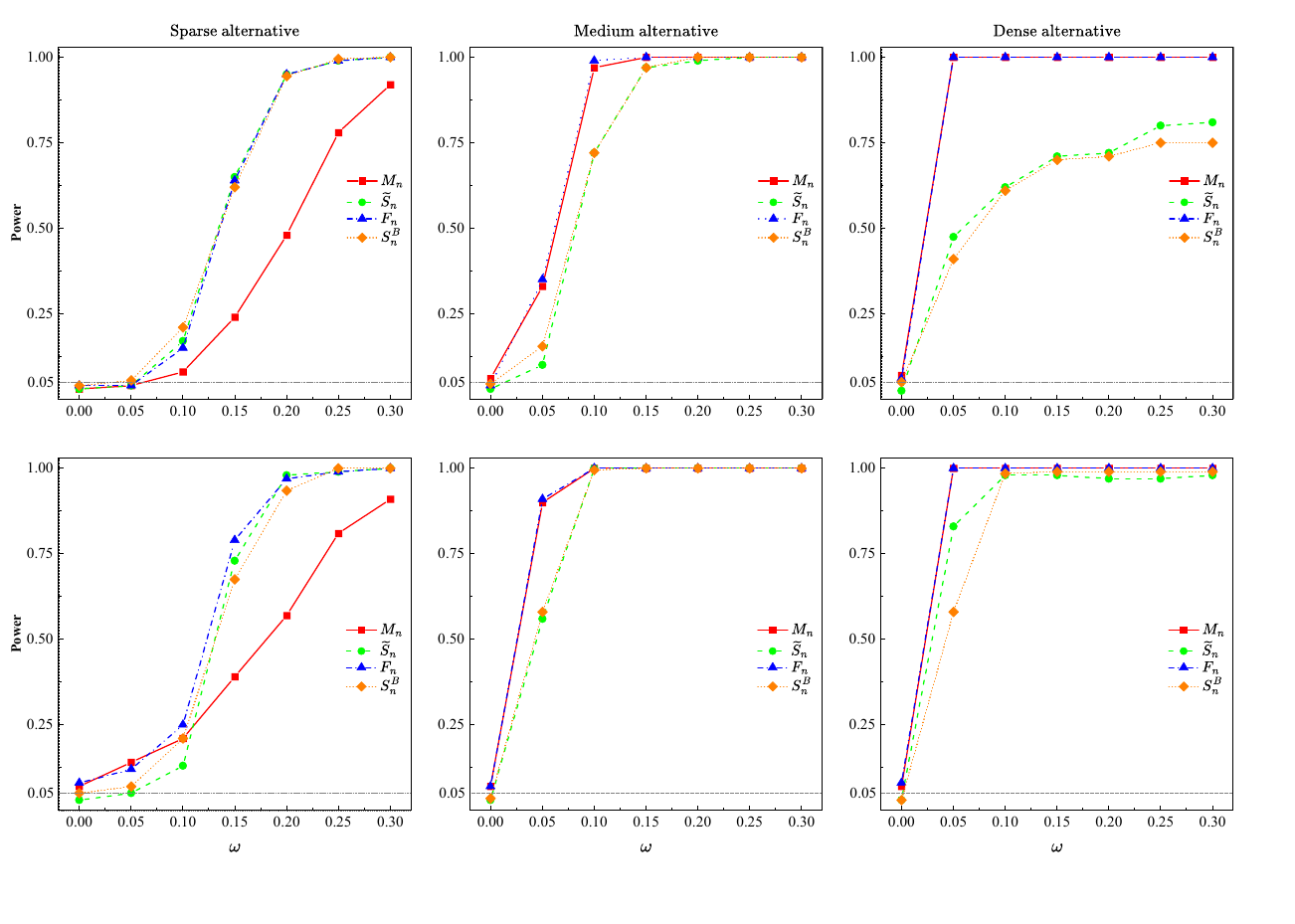}
		\caption{Power curves   with  $(n,p)=(200,100)$ and $\boldsymbol{\beta}^*=\omega\ast\left(\mathbf{1}_{s},\mathbf{0}_{p-s}\right)^{\top}$. 
			The  $M_n$, $\tilde{S}_n$ and   $F_n$  correspond to the results derived from the quadratic-type test,   maximum-type test and Fisher adaptive test  proposed  in this paper,  respectively,  while $S_n^B$ represents the maximum-type test from  \cite{beyhum2023tuning}. 
			The first row represent the results obtained with the covariance matrix $\boldsymbol{I}_p$, while the second row corresponds to the results with covariance matrix $\boldsymbol{\Sigma}_{\boldsymbol{u}}^{\text{AR}}$.
			The first column reports results for the fully sparse case ($s=1$),  the second column for the moderately dense case  ($s=10$), and the third column for the fully dense case ($s=p$).  } \label{p_100_sfix}
	\end{figure}

\begin{figure}[H]%
		\centering
		\includegraphics[width=1.1\textwidth]{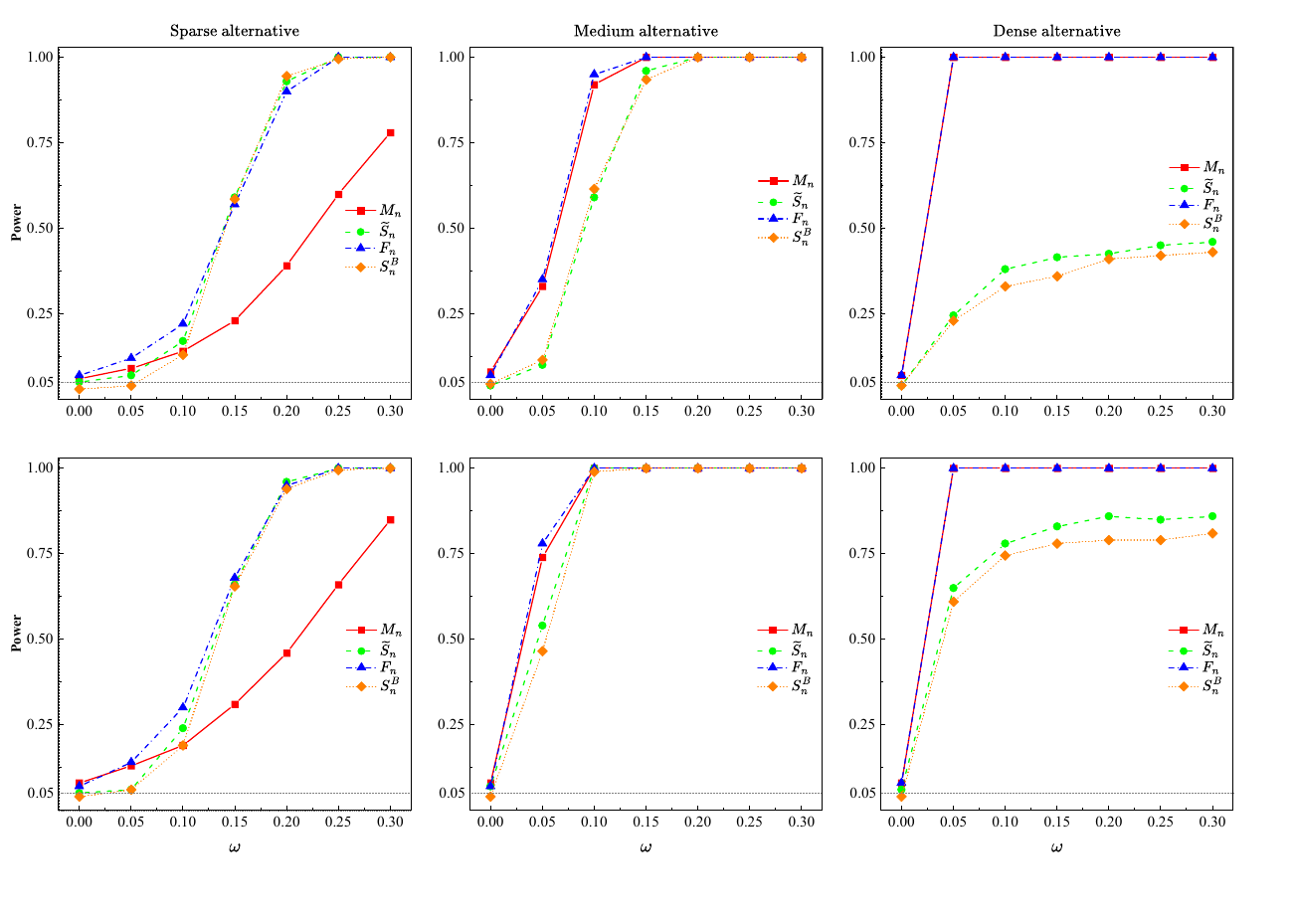}
		\caption{Power curves   with  $(n,p)=(200,200)$ and $\boldsymbol{\beta}^*=\omega\ast\left(\mathbf{1}_{s},\mathbf{0}_{p-s}\right)^{\top}$. 
			The  $M_n$, $\tilde{S}_n$ and   $F_n$  correspond to the results derived from the quadratic-type test,   maximum-type test and Fisher adaptive test  proposed  in this paper,  respectively,  while $S_n^B$ represents the maximum-type test from  \cite{beyhum2023tuning}.  
			The first row represent the results obtained with the covariance matrix $\boldsymbol{I}_p$, while the second row corresponds to the results with covariance matrix $\boldsymbol{\Sigma}_{\boldsymbol{u}}^{\text{AR}}$.
			The first column reports results for the fully sparse case ($s=1$),  the second column for the moderately dense case  ($s=10$), and the third column for the fully dense case ($s=p$).  }\label{p_200_sfix}
	\end{figure}

		\begin{figure}[H]%
		\centering
		\includegraphics[width=1.1\textwidth]{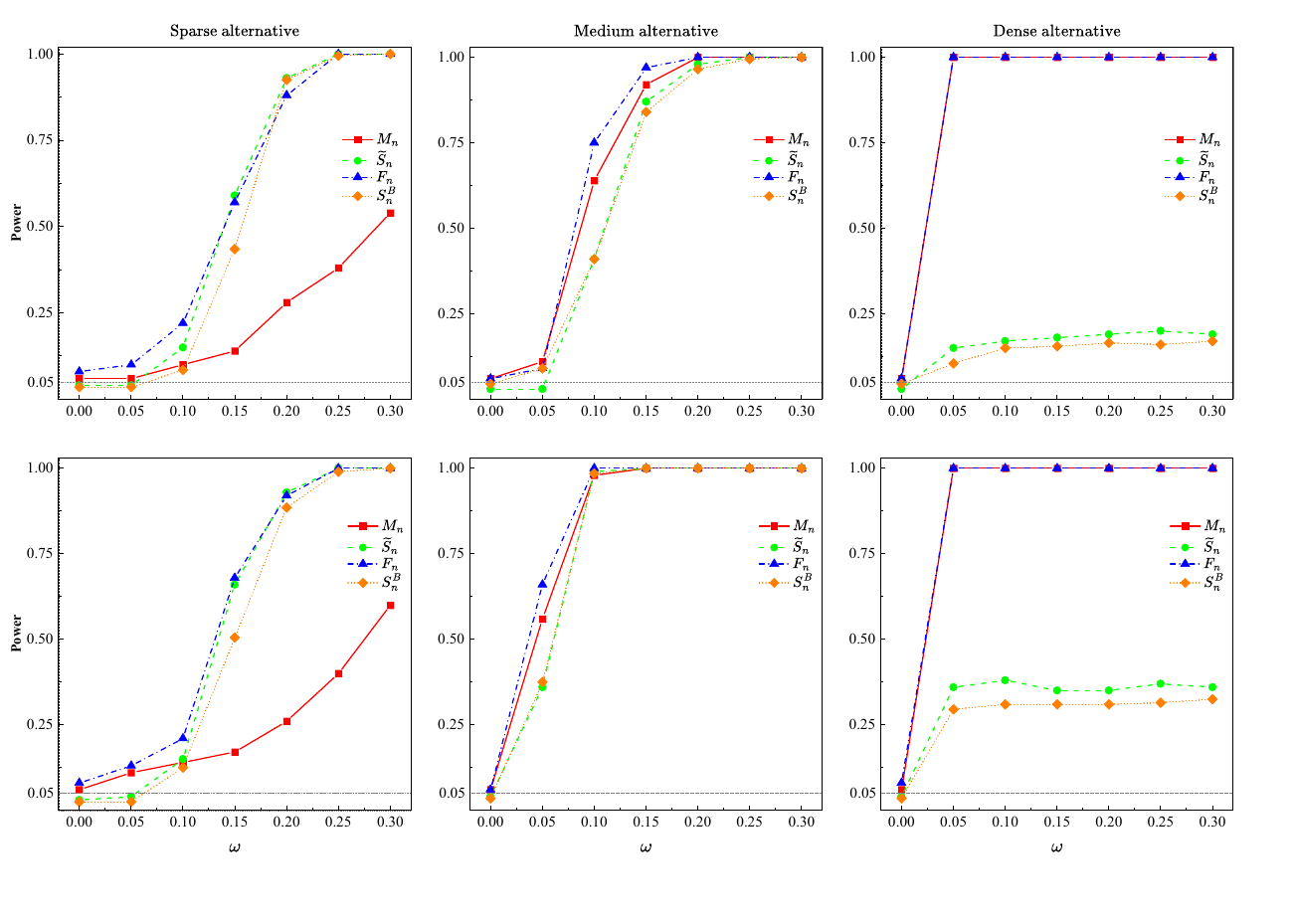}
		\caption{Power curves   with  $(n,p)=(200,500)$ and $\boldsymbol{\beta}^*=\omega\ast\left(\mathbf{1}_{s},\mathbf{0}_{p-s}\right)^{\top}$. 
			The  $M_n$, $\tilde{S}_n$ and   $F_n$  correspond to the results derived from the quadratic-type test,   maximum-type test and Fisher adaptive test  proposed  in this paper,  respectively,  while $S_n^B$ represents the maximum-type test from  \cite{beyhum2023tuning}. 
			The first row represent the results obtained with the covariance matrix $\boldsymbol{I}_p$, while the second row corresponds to the results with covariance matrix $\boldsymbol{\Sigma}_{\boldsymbol{u}}^{\text{AR}}$.
			The first column reports results for the fully sparse case ($s=1$),  the second column for the moderately dense case  ($s=10$), and the third column for the fully dense case ($s=p$). }\label{p_500_sfix}
	\end{figure}

	\begin{figure}[H]%
		\centering
		\includegraphics[width=1.1\textwidth]{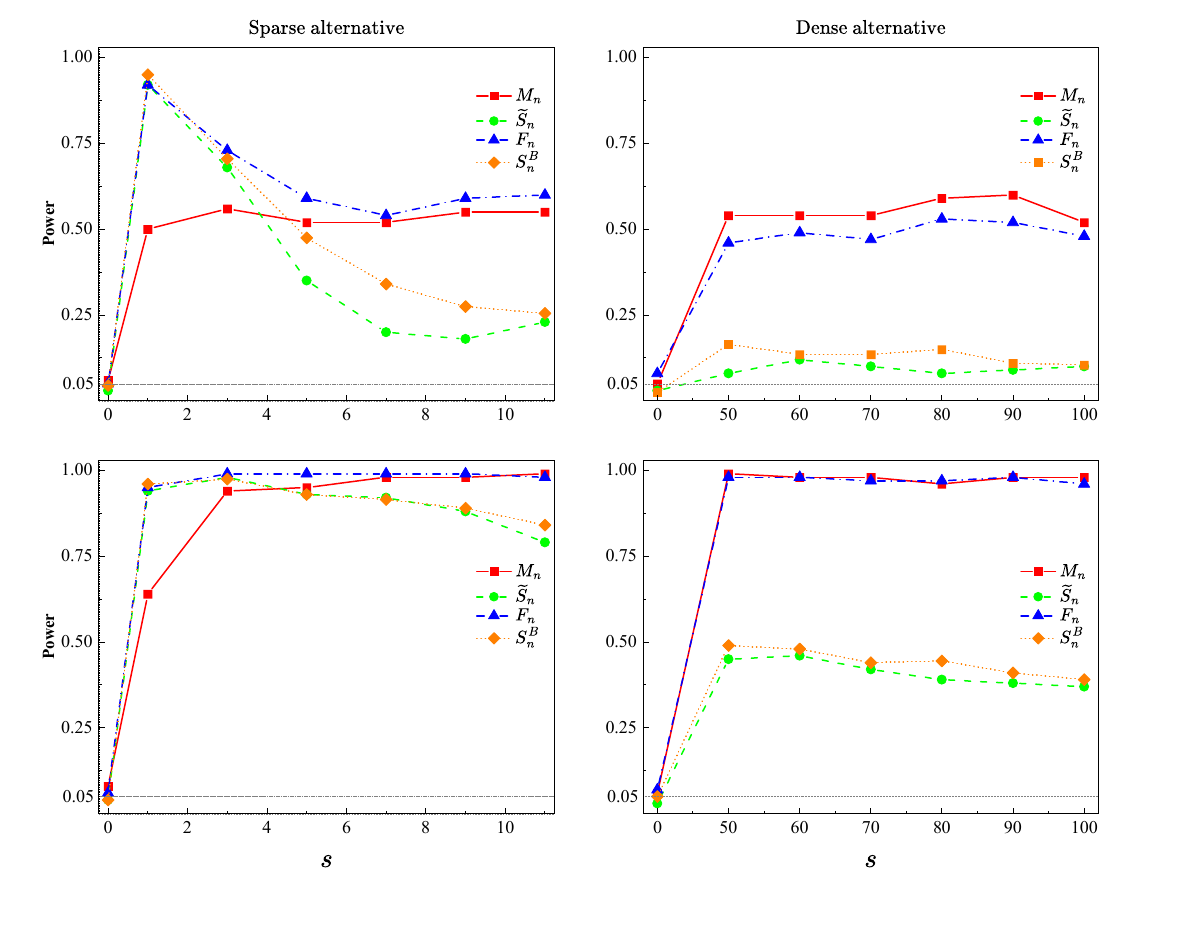}
		\caption{Power curves   with  $(n,p)=(200,100)$ and $\|\boldsymbol{\beta}^*\|_2=0.2$. 
			The  $M_n$, $\tilde{S}_n$ and   $F_n$  correspond to the results derived from the quadratic-type test,   maximum-type test and Fisher adaptive test  proposed  in this paper,  respectively,  while $S_n^B$ represents the maximum-type test from  \cite{beyhum2023tuning}. 
			The first row represent the results obtained with the covariance matrix $\boldsymbol{I}_p$, while the second row corresponds to the results with covariance matrix $\boldsymbol{\Sigma}_{\boldsymbol{u}}^{\text{AR}}$.
			The first column shows the results under the sparse case (\(s = 1, 3, 5, 7, 9, 11\)), while  the second column exhibits the results under the dense case (\(s = 50\%p, 60\%p, 70\%p, 80\%p, 90\%p, 100\%p\)).  } \label{p_100_fig}
	\end{figure}

	\begin{figure}[H]%
		\centering
		\includegraphics[width=1.1\textwidth]{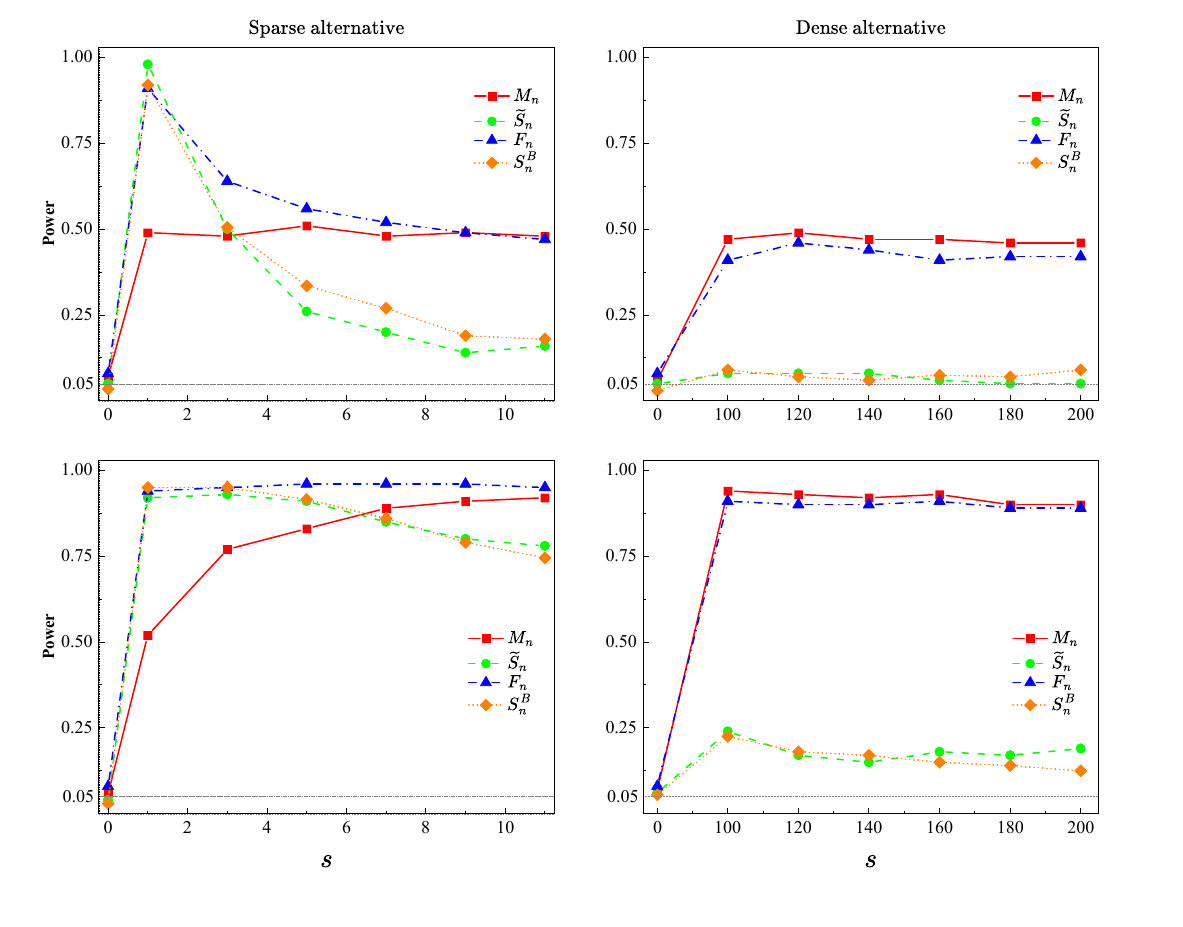}
		\caption{Power curves   with  $(n,p)=(200,200)$ and $\|\boldsymbol{\beta}^*\|_2=0.2$. 
			The  $M_n$, $\tilde{S}_n$ and   $F_n$  correspond to the results derived from the quadratic-type test,   maximum-type test and Fisher adaptive test  proposed  in this paper,  respectively,  while $S_n^B$ represents the maximum-type test from  \cite{beyhum2023tuning}. 
			The first row represent the results obtained with the covariance matrix $\boldsymbol{I}_p$, while the second row corresponds to the results with covariance matrix $\boldsymbol{\Sigma}_{\boldsymbol{u}}^{\text{AR}}$.
			The first column shows the results under the sparse case (\(s = 1, 3, 5, 7, 9, 11\)), while  the second column exhibits the results under the dense case (\(s = 50\%p, 60\%p, 70\%p, 80\%p, 90\%p, 100\%p\)).   }\label{p_200_fig}
	\end{figure}

	\begin{figure}[H]%
		\centering
		\includegraphics[width=1.1\textwidth]{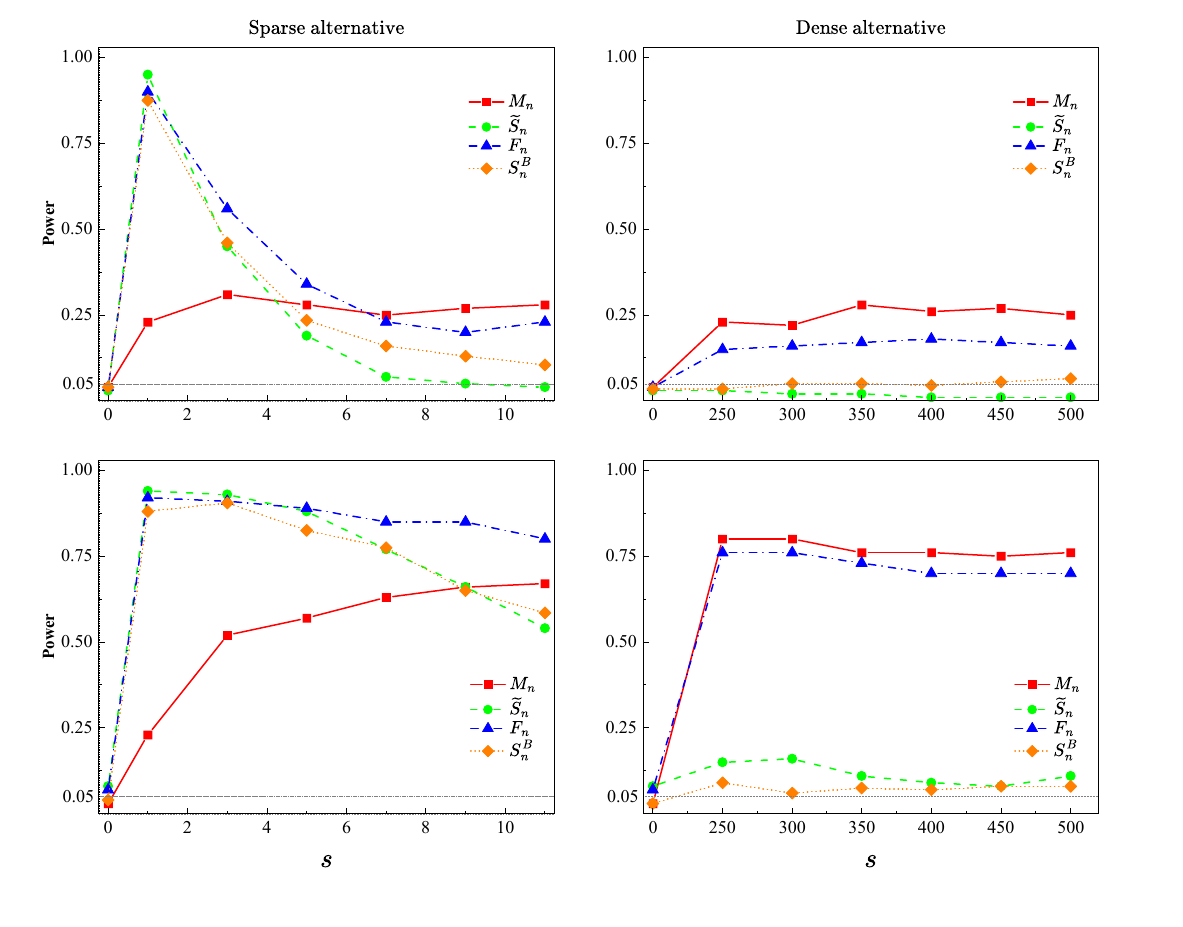}
		\caption{Power curves   with  $(n,p)=(200,500)$ and $\|\boldsymbol{\beta}^*\|_2=0.2$. 
			The  $M_n$, $\tilde{S}_n$ and   $F_n$  correspond to the results derived from the quadratic-type test,   maximum-type test and Fisher adaptive test  proposed  in this paper,  respectively,  while $S_n^B$ represents the maximum-type test from  \cite{beyhum2023tuning}. 
			The first row represent the results obtained with the covariance matrix $\boldsymbol{I}_p$, while the second row corresponds to the results with covariance matrix $\boldsymbol{\Sigma}_{\boldsymbol{u}}^{\text{AR}}$.
			The first column shows the results under the sparse case (\(s = 1, 3, 5, 7, 9, 11\)), while  the second column exhibits the results under the dense case (\(s = 50\%p, 60\%p, 70\%p, 80\%p, 90\%p, 100\%p\)).  }\label{p_500_fig}
	\end{figure}

\end{document}